%% file: main.tex
\documentclass[sigplan, nonacm]{acmart}
\setcopyright{none}

\usepackage[T1]{fontenc}
\usepackage[english]{babel}
\usepackage[utf8]{inputenc}
\usepackage{amsfonts,amsmath} 
\usepackage[all,arc]{xy}
\usepackage{mathrsfs}
\usepackage{comment}
\usepackage{graphicx} 
\usepackage{commath}
\usepackage{afterpage}
\usepackage{xcolor}
\usepackage{paralist}
\usepackage{caption}
\usepackage{subcaption}
\usepackage{natbib}
\usepackage{fancybox}
\usepackage{array}
\usepackage{booktabs}
\usepackage{pifont}
\newcommand{\xmark}{\ding{55}}%
\usepackage{microtype}

\usepackage[disable]{todonotes} 
\newcommand{\JEDI}[1]{\todo[color=cyan!5,inline]{#1}}
\newcommand{\TODO}[1]{\todo[color=red!10,inline]{TODO: #1}}

\usepackage[T1]{fontenc}

\usepackage[ruled, linesnumbered,vlined]{algorithm2e}

\theoremstyle{remark}

\usepackage{listings}
\usepackage{relsize}
\lstdefinestyle{default}{%
  basicstyle=\ttfamily,%
  commentstyle=\sl,%
  keywordstyle=\bf\relsize{-0.5},%
  columns=fullflexible,%
  keepspaces=true,%
  mathescape,%
  escapechar=\#%
}
\lstdefinestyle{number}{%
  numbers=left,%
  numberstyle=\scriptsize\em,%
  xleftmargin=2em%
}
\newcommand{\mylstnumber}{$\loc_{\arabic{lstnumber}}$}
\renewcommand{\thelstnumber}{\mylstnumber}
\makeatletter
\newcount\lst@savelstnumber
\newcommand{\lstnonum}{%
  \global\lst@savelstnumber\c@lstnumber%
  \gdef\thelstnumber{}%
}
\newcommand{\lststopn}{%
  \lstnonum%
}
\newcommand{\lststartn}{%
  \global\c@lstnumber\lst@savelstnumber%
  \gdef\thelstnumber{\mylstnumber}
}
\newcommand{\lstbeginn}{%
  \global\c@lstnumber-1%
  \gdef\thelstnumber{\mylstnumber}%
}
\newcommand{\lstoverriden}[1]{%
  \lstnonum%
  \gdef\thelstnumber{#1}%
}
\makeatother

\lstdefinelanguage{DiffC}[]{C++}{
  morekeywords={assume}
}
\definecolor{lightredbg}{rgb}{1.0,0.93,0.93}
\definecolor{lightgreenbg}{rgb}{0.92,1.0,0.92}
\definecolor{lightyellowbg}{rgb}{1.0,1.0,0.88}
\lstnewenvironment{lstcont}[1][]
 {\lstset{aboveskip=-\medskipamount,firstnumber=last,#1}}
 {}
\lstnewenvironment{lstdiffplus}[1][]
 {\lstset{aboveskip=-\medskipamount,firstnumber=last,backgroundcolor=\color{lightgreenbg},#1}}
 {}
\lstnewenvironment{lstdiffminus}[1][]
 {\lstset{aboveskip=-\medskipamount,firstnumber=last,backgroundcolor=\color{lightredbg},#1}}
 {}
\lstnewenvironment{lstcostmodel}[1][]
 {\lstset{aboveskip=-\medskipamount,firstnumber=last,backgroundcolor=\color{lightyellowbg},#1}}
 {}
\newcommand{\diffplus}[1]{\colorbox{lightgreenbg}{#1}}
\newcommand{\diffminus}[1]{\colorbox{lightredbg}{#1}}
\newcommand{\costmodel}[1]{\colorbox{lightyellowbg}{#1}}

\usepackage{tikz}
\usepackage{tikz,pgffor}
\usepackage{changepage}
\usetikzlibrary{arrows}
\usetikzlibrary{shapes}
\usetikzlibrary{calc}
\usetikzlibrary{automata}
\usetikzlibrary{positioning}
\usetikzlibrary{angles}

\tikzstyle{ang}=[regular polygon, regular polygon sides = 3,draw,inner sep=0pt,minimum size=6mm, yshift = -0.75 mm]
\tikzstyle{dem}=[shape=diamond,draw,inner sep=0pt,minimum size=6mm]
\tikzstyle{ran}=[shape=circle,draw,inner sep=0pt,minimum size=6mm]
\tikzstyle{det}=[shape=rectangle,draw,inner sep=0pt,minimum size=5mm]
\tikzstyle{tran}=[draw,->,>=stealth, rounded corners]

\newcommand{\vars}{V}

\newcommand{\locs}{\mathit{L}}
\newcommand{\transys}{\mathcal{T}}
\newcommand{\loc}{\ell}
\newcommand{\locinit}{\ensuremath{\loc_0}}
\newcommand{\varinit}{\mathbf{x}_0}
\newcommand{\initvars}{\Theta_0}

\newcommand{\cost}{\mathtt{cost}} 
\newcommand{\Cost}{\mathit{Cost}}
\newcommand{\CostSup}{\mathit{CostSup}}
\newcommand{\CostInf}{\mathit{CostInf}}
\newcommand{\trans}{\rightarrow}
\newcommand{\lout}{\fmttinylbl{out}}
\newcommand{\locterm}{\ensuremath{\loc_{\lout}}}
\newcommand{\Run}{\mathit{Run}}
\newcommand{\Guard}{\mathit{G}}
\newcommand{\Update}{\mathit{Up}}

\newtheorem{manualtheoreminner}{\bf Theorem}
\newenvironment{manualtheorem}[1]{%
	\renewcommand\themanualtheoreminner{#1}%
	\manualtheoreminner
}{\endmanualtheoreminner}

\newcommand{\fmttinylbl}[1]{\ensuremath{\text{\rm\tiny #1}}}
\newcommand{\fmtsublbl}[2]{\ensuremath{#2^{\fmttinylbl{#1}}}}
\newcommand{\oldsub}[1]{\fmtsublbl{old}{#1}}
\newcommand{\newsub}[1]{\fmtsublbl{new}{#1}}
\newcommand{\oldcost}{\oldsub{\cost}}
\newcommand{\newcost}{\newsub{\cost}}

\newcommand{\threshold}{\ensuremath{t}}

\newcommand{\potential}{\phi}
\newcommand{\antipotential}{\chi}

\newcommand{\myparagraph}[1]{\par\smallskip\noindent\emph{#1}}

\begin{document}

\title{Differential Cost Analysis with Simultaneous Potentials and Anti-potentials}



\author{\DJ or\dj e \v{Z}ikeli\'c}\authornote{This paper describes work performed in part while \DJ or\dj e \v{Z}ikeli\'c was an Applied Scientist Intern at Amazon.}
\affiliation{IST Austria, Austria}
\email{dzikelic@ist.ac.at}

\author{Bor-Yuh Evan Chang}\authornote{Bor-Yuh Evan Chang holds concurrent appointments at the University of Colorado Boulder and as an Amazon Scholar. This paper describes work performed at Amazon and is not associated with CU Boulder.}
\affiliation{Amazon, USA}
\affiliation{University of Colorado Boulder, USA}
\email{byec@amazon.com}

\author{Pauline Bolignano}
\affiliation{Amazon, UK}
\email{pln@amazon.com}

\author{Franco Raimondi}\authornote{Franco Raimondi holds concurrent appointments at Middlesex University and as an Amazon Scholar. This paper describes work performed at Amazon and is not associated with Middlesex.}
\affiliation{Amazon, UK}
\affiliation{Middlesex University, UK}
\email{frai@amazon.com}


\begin{abstract}
We present a novel approach to differential cost analysis that, given a program revision, attempts to statically bound the difference in resource usage, or cost, between the two program versions.
Differential cost analysis is particularly interesting because of the many compelling applications for it, such as detecting resource-use regressions at code-review time or proving the absence of certain side-channel vulnerabilities.
One prior approach to differential cost analysis is to apply relational reasoning that conceptually constructs a product program on which one can over-approximate the difference in costs between the two program versions.
However, a significant challenge in any relational approach is effectively aligning the program versions to get precise results.
In this paper, our key insight is that we can avoid the need for and the limitations of program alignment if, instead, we bound the difference of two cost-bound summaries rather than directly bounding the concrete cost difference.
In particular, our method computes a threshold value for the maximal difference in cost between two program versions \emph{simultaneously} using two kinds of cost-bound summaries---a potential function that evaluates to an upper bound for the cost incurred in the first program and an \emph{anti-potential} function that evaluates to a lower bound for the cost incurred in the second.
Our method has a number of desirable properties: it can be fully automated, it allows optimizing the threshold value on relative cost, it is suitable for programs that are not syntactically similar, and it supports non-determinism.
We have evaluated an implementation of our approach on a number of program pairs collected from the literature, and we find that our method computes tight threshold values on relative cost in most examples.


	
\end{abstract}

\maketitle

\input{intro}

\input{overview}
\input{prelims}

\input{potentialfunctions}

\input{algorithm}

\input{exp}

\input{tightness}

\input{relatedwork}

\section*{Acknowledgements}

We thank Shaun Willows, Thomas Lugnet, and the Living Room Application Vending team for suggesting threshold bounds as a developer-friendly way to interact with a differential cost analyzer, and we thank Jim Christy, Daniel Schoepe, and the Prime Video Automated Reasoning team for their support and helpful suggestions throughout the project.
We also thank Michael Emmi for feedback on an earlier version of this paper.
And finally, we thank the anonymous reviewers for their useful feedback and Aws Albarghouthi for shepherding the final version of the paper.
\DJ or\dj e \v{Z}ikeli\'c was also partially supported by ERC CoG 863818 (FoRM-SMArt).

\bibliography{diffcostanalysis}

\clearpage
\appendix

\input{app}

\end{document}

%% file: intro.tex
\section{Introduction}\label{sec:intro}

\JEDI{Problem: We want to compute the difference in costs between two program versions.}
We consider the problem of statically bounding the \emph{difference} in resource usage (i.e., \emph{cost}) between two program versions. 
In particular, for two program versions and a set of inputs, the goal of differential cost analysis is to compute a {\em threshold} value $\threshold$ that bounds the maximal difference in cost usage between the two programs.
That is, let numeric-valued variables $\oldcost$ and $\newcost$ model the resource usage of the old and new program versions, respectively, then we want to prove the \emph{differential threshold bound} assertion:
\[
	\newcost - \oldcost \leq \threshold
\]
on termination of the program versions when given the same input (for all inputs).
Notice that our notion of cost is quite generic, and it encompasses metrics such as run time, memory usage, the number of object allocations or the number of thread allocations, etc.

\JEDI{Why Important: Want to detect regressions at code review time in CI systems.}
This static analysis problem has many practical applications.
For instance, in software development, programs are often modified and extended with new features.
A program revision might lead to unacceptable jumps in cost usage.
For performance critical software, it is crucial to detect such undesired performance regressions prior to releasing the software to production.
With static analysis seeing increasing industrial adoption for verification or bug finding (e.g., in continuous integration pipelines or backing automated code reviews), differential cost analysis is a critical tool to catch potential performance regressions early in development.
%

\JEDI{Why Hard: One clear way is to bound the difference in costs between the two versions is to use relational analysis. But one limitation or challenge of applying this approach is that it requires program-version alignment. But given an alignment, you get a product program, and you can apply whatever cost analysis approach you want on the product program, such as those based on potential functions. That's what prior work on Relational Cost Analysis does.}
One clear way to address the differential cost analysis problem is to apply the \emph{relational approach}~\cite{DBLP:conf/popl/Benton04,DBLP:journals/mscs/BartheDR11} where one
reasons about the so-called product or relational state of two program versions.
%
There are several recent works on relational cost analysis~\cite{CicekBG0H17,RadicekBG0Z18,CicekQBG019} that do essentially this. These works consider functional programs and use relational type systems for reasoning about and bounding the difference in cost between two programs; the recent work of \citet{QuG019} additionally allows array-manipulating programs with a relational type-and-effect system.
While these works present significant advances, a substantial difficulty behind relational reasoning in general is effectively aligning the two programs, as parts that cannot be aligned fall back to less precise unary reasoning. 

\JEDI{Contribution: Our insight is that we can instead take the difference of cost bounds of the two program versions, rather than bounding the difference of the product program, eliminating the need and limitation of version alignment. Of course, it is tricky to do this soundly and precisely---we need to be careful about when to use upper and when to use lower bounds. We need an approach to compute lower bounds. That's where we introduce a lower-bound analogue of potential functions called anti-potential functions. And we need to do simultaneous reasoning to be precise, which we can do by aligning the entry point.}

Our insight in this work is that if we instead take the difference of cost bounds of the two program versions to compare with the differential threshold, then we can sidestep the need for and the limitations of program alignment. Of course, doing this na\"ively could be problematic either for soundness or precision. For soundness, we must carefully use both upper and lower bounds on $\cost$ (i.e., take the difference between the upper bound of $\newcost$ and the lower bound of $\oldcost$). Lower bounds may be a challenge to get, as the existing methods and tools for cost analysis primarily focus on computing upper bounds. For precision, these cost bounds cannot be computed completely independently on the two program versions, as we seek to prove tight threshold bounds when each program version is given the same input.

In this paper, we propose a new method for differential cost analysis in numerical imperative programs with polynomial arithmetic and with non-determinism.
Our method uses potential functions from amortized analysis~\cite{tarjan1985amortized} to reason about costs incurred in individual programs. Potential functions are a well-known method for computing upper bounds on the cost incurred in a single program~\cite{HoffmannAH11,HoffmannH10,HoffmannDW17,Carbonneaux0S15,0002AH12}.
Intuitively, a potential function assigns a ``valuation'' to a program state that is sufficient to ``pay'' for the resource use along all paths to a terminating state.
What we do in this work is to also use a lower-bound analogue---that we call \emph{anti-potential} functions---to reason about the relative cost between two program versions.
An anti-potential function instead assigns a ``valuation'' to a program state that is \emph{insufficient} to ``pay'' for the resource use along all paths to a terminating state.
While we have drawn inspiration from lower-bound analogues in other domains~\cite{FrohnNBG20,Wang0GCQS19,NgoDFH17}, the key contribution here is computing a differential threshold value on the maximal difference in cost between two programs {\em simultaneously} with potential and anti-potential functions---one that provides an upper bound on the cost incurred in the new version and the other that provides a lower bound on the cost incurred in the old for the \emph{same inputs}.
The simultaneous computation is done by employing a constraint solving-based approach, which collects the necessary constraints on potential and anti-potential functions to serve as upper and lower bounds on incurred cost in the program versions, as well as the differential cost constraint.

\JEDI{What Follows: We have a differential cost analysis approach with several nice properties: (1) we do not need a program version alignment, (2) we make few assumptions about the programming language (can be imperative, non-deterministic), though soundness is guaranteed only if program versions terminate and have some precision guarantees only if the program versions are deterministic.}

The constraint solving-based approach allows our method to provide several key properties: (1)~our method can be fully automated, (2)~the computation of the threshold value and witnessing potential functions proceed by reduction to linear programming, hence it allows efficient {\em optimization} of the threshold value by introducing a minimization objective in the linear program, (3)~since our method does not depend on syntactic alignment of programs and performs relational reasoning only on the level of inputs, it is suitable for programs that are not syntactically similar, and (4)~our method supports non-determinism in the programming language (which can be hard to support with program alignment).

Finally, while we focus on proving differential threshold bound assertions for two program versions and on optimizing the threshold value, we also show that our method can be used to prove any given {\em symbolic polynomial bound} in terms of program inputs on the difference in cost usage. The reason to focus on concrete threshold values is that they can be optimized, since the real numbers form a well-ordered set. In contrast, it is not clear what would be a correct criterion for optimizing polynomial bounds on a set of inputs, which is the reason why we are only able to prove a given symbolic polynomial bound on the difference in cost usage.

\myparagraph{Contributions.} Our contributions are as follows:
\begin{itemize}
	\item We present a new method for differential cost analysis that uses potential and anti-potential functions to reason about relative incurred cost (Section~\ref{sec:potentialsection}).

	\item We give an algorithm for deriving potential and anti-potential functions simultaneously with a threshold value
	that verifies the differential threshold bound assertion
	in imperative numerical programs with polynomial arithmetic and non-determinism (Section~\ref{sec:algorithm}).

	\item Our experimental evaluation demonstrates the ability of our method to compute \emph{tight} threshold values for differential cost analysis (Section~\ref{sec:experiments}).
\end{itemize}
Also as a consequence of our approach, we show that the potential and anti-potential method can be adapted to provide a way for computing tight cost bounds on single programs, that is, bounds with precision guarantees (see Section~\ref{sec:tight}).

%% file: overview.tex
\section{Overview}\label{sec:overview}

To illustrate our approach and new concepts, we consider the program pair shown in Fig.~\ref{fig:running}, which will serve as our running example throughout this paper.
%
We assume that each program has a special variable $\cost$ that is initialized to $0$ and is updated whenever cost is incurred in the program. It is defined by the underlying cost model that could track program run time, memory usage, or any other quantitative property of interest. Cost may take both positive and negative values. In order for the total cost of a program run to be well-defined, we assume that our programs are {\em terminating}. We elaborate on the need for this assumption in Section~\ref{sec:prelims}.

\JEDI{I modified the example so that programs differ both in the loop ordering and in the version of f that they call. The former is necessary so that the programs cannot be syntactically aligned. After that I explain that, even if the loop ordering is the same, if f is called through an API then only a fully automated method can detect the cost regression. Hence, this example motivates (1)~a need for a method for differential cost analysis that does not assume syntactic alignment, and (2)~a need for a fully automated method.}




\newcommand{\arrA}{\ensuremath{\mathtt{A}}}
\newcommand{\arrB}{\ensuremath{\mathtt{B}}}
\newcommand{\lenA}{\ensuremath{\mathtt{lenA}}}
\newcommand{\lenB}{\ensuremath{\mathtt{lenB}}}
\newcommand{\idxi}{\ensuremath{\mathtt{i}}}
\newcommand{\idxj}{\ensuremath{\mathtt{j}}}
\newcommand{\assertbox}[1]{\ovalbox{\vphantom{$\oldsub{\antipotential}\newsub{\potential}$}#1}}
\newsavebox{\SBoxLocterm}\sbox{\SBoxLocterm}{\scriptsize\em\locterm} 
\newsavebox{\SBoxMOld}
\begin{lrbox}{\SBoxMOld}
\begin{minipage}{0.49\textwidth}
\lstnonum\begin{lstlisting}[style=number]
void join(int A[], int lenA, int B[], int lenB) {
  assume($1 \leq\;$lenA$\;\leq 100 \,\land\, 1 \leq \;$lenB$\;\leq 100$);
\end{lstlisting}
\begin{lstcostmodel}[style=number]
  int cost = 0;
\end{lstcostmodel}
\lststartn\begin{lstcont}[style=number,firstnumber=0]
  #\assertbox{$\oldsub{\antipotential}\colon$lenA$\,\cdot\,$lenB}\label{loc-init}\lststopn#
  for (int i = 0;$\lststartn$
  #\assertbox{$\oldsub{\antipotential}\colon ($lenA$\,-\,$i$) \,\cdot\, $lenB}\label{loc-outer-init}\lststopn#
      i < #\diffminus{lenA}#; i++) {
    for (int j = 0;$\lststartn$
    #\assertbox{$\oldsub{\antipotential}\colon ($lenA$\,-\,$i$) \,\cdot\, $lenB$ \,-\, $j}\label{loc-inner-init}\lststopn#
        j < #\diffminus{lenB}#; j++) {$\lststartn$
      #\assertbox{$\oldsub{\antipotential}\colon ($lenA$\,-\,$i$) \,\cdot\, $lenB$ \,-\, $j}\label{loc-inner-body}\lststopn#
      f(#\diffminus{A[i]}#, #\diffminus{B[j]}##\costmodel{, cost}#);
    }
  }#\lstoverriden{\usebox{\SBoxLocterm}}#
  #\assertbox{$\oldsub{\antipotential}\colon 0$}\label{loc-term}\lststopn#
}
\end{lstcont}
\begin{lstlisting}[style=number]
void f(int a, int b#\costmodel{, int \&cost}#) {
\end{lstlisting}
\begin{lstdiffminus}[style=number]
  $\ldots$
\end{lstdiffminus}
\begin{lstcostmodel}[style=number]
  cost = cost + 1;
\end{lstcostmodel}
\begin{lstcont}[style=number]
}
\end{lstcont}
\end{minipage}
\end{lrbox}
\newsavebox{\SBoxMNew}
\begin{lrbox}{\SBoxMNew}
\begin{minipage}{0.45\textwidth}
\begin{lstlisting}
void join(int A[], int lenA, int B[], int lenB) {
  assume($1 \leq \lenA \leq 100 \,\land\, 1 \leq lenB \leq 100$);
\end{lstlisting}
\begin{lstcostmodel}
  int cost = 0;
\end{lstcostmodel}
\begin{lstcont}
  #\assertbox{$\newsub{\potential}\colon 2 \,\cdot\, $lenB$\,\cdot\,$lenA}#
  for (int i = 0;
  #\assertbox{$\newsub{\potential}\colon  2\,\cdot\,($lenB$\,-\,$i$) \,\cdot\, $lenA$)$}#
      i < #\diffplus{lenB}#; i++) {
    for (int j = 0;
    #\assertbox{$\newsub{\potential}\colon 2\cdot(($lenB$\,-\,$i$) \,\cdot\, $lenA$ \,-\, $j$)$}#
        j < #\diffplus{lenA}#; j++) {
      #\assertbox{$\newsub{\potential}\colon 2\cdot(($lenB$\,-\,$i$) \,\cdot\, $lenA$ \,-\, $j$)$}#
      f(#\diffplus{A[j]}#, #\diffplus{B[i]}##\costmodel{, cost}#);
    }
  }
  #\assertbox{$\newsub{\potential}\colon 0$}#
}
\end{lstcont}
\begin{lstlisting}
void f(int a, int b#\costmodel{, int \&cost}#) {
\end{lstlisting}
\begin{lstdiffplus}
  $\ldots$
\end{lstdiffplus}
\begin{lstcostmodel}
  cost = cost + 2;
\end{lstcostmodel}
\begin{lstcont}
}
\end{lstcont}
\end{minipage}
\end{lrbox}
\begin{figure*}[tb]
\usebox{\SBoxMOld}
\hfill
\usebox{\SBoxMNew}
\caption{An example revision of a procedure \lstinline{join} affecting its cost with the old version on the left and new one on the right.
One can see the \lstinline{join} procedure as representing a common join over two sequences (e.g., arrays, lists, collections via iterators) with an operator \lstinline{f} having some cost per pair of elements.
For subsequent discussion, we label some important program points in \lstinline{join} with \locinit--\locterm{} (conceptually, locations in \lstinline{join}'s control-flow graph).
To model the cost of code, it is standard to introduce a $\cost$ variable, which we can see as ghost state.
For improved readability, we highlight the ghost code for updating $\cost$ in \costmodel{yellow}.
The goal of differential cost analysis on \lstinline{join} is to compute a bound on the difference between $\cost$ in the two versions on termination at program point~\locterm.
There are two conceptual changes in this revision.
First, in \lstinline{join} itself, the loops are interchanged, which absent any other changes should not affect the total cost.
We highlight deletions in \diffminus{red} and additions in \diffplus{green}.
Second, there is some revision in the \lstinline{f} operation that changes the cost per pair from 1 to 2 on every input.
At program points \locinit--\locterm, we show potentials $\potential$ and anti-potentials $\antipotential$ that enable proving that the relative cost is at most $\lenA \cdot \lenB$ (which is bounded by 10,000 when assuming both $\lenA$ and $\lenB$ are bounded by 100).
}
\label{fig:running}
\end{figure*}

\begin{example}[An example revision]\label{ex:running}
	Fig.~\ref{fig:running} shows a revision to a procedure \lstinline{join} that calls an operator \lstinline{f} to which a revision was also made.
	Both versions take two integer arrays $\arrA$ and $\arrB$ and their lengths $\lenA$ and $\lenB$ as inputs, consist of a nested loop that iterates through both arrays, and on each inner loop iteration call the operator \lstinline{f}. However, the orderings in which the two versions iterate through the arrays differ. Hence, the difference introduced by the revision is reflected in the change of ordering of loops and in the fact that \lstinline{f} is modified in a way that changes its cost.
	
	Due to the different loop ordering, two versions of \lstinline{join} cannot be syntactically aligned, making it difficult to apply the relational approach.
	On the other hand, even the version of this example without the loop interchange shows a pattern that is very easy to imagine in practice, and that may introduce unexpected jumps in cost usage.
	For instance, suppose that \lstinline{f} is called through an interface so the programmer that implements \lstinline{join} may not have the source code of \lstinline{f} immediately available to them. In such situations, manual detection and quantification of potential increase in cost usage resulting from the revision is not possible, and an automated differential cost analysis is necessary in order to warn the programmer about the potential performance regression. Thus, this example illustrates the challenges in differential cost analysis that our work aims to address.
\end{example}



\JEDI{Informal definition of the threshold problem. Change: RCost prolem -> DiffCost problem}

\myparagraph{The differential threshold problem.} To quantify the change in cost between two programs, we consider the problem of computing a {\em threshold} value that bounds the maximal difference in their cost usage when given the same input. 
We refer to this as the {\em differential cost analysis problem} (or \emph{DiffCost} for short).
A key feature of our approach is that we not only compute a threshold for the DiffCost problem but also {\em optimize} it to compute as tight of a threshold as possible.

\JEDI{Recalling PFs and introducing anti-PFs, plus motivating anti-PFs as a dual notion. The paragraph is followed by an example on PFs and anti-PFs.}

\myparagraph{Potential and anti-potential functions.} To reason about costs incurred in individual programs, we use potential functions (PFs)~\cite{tarjan1985amortized}. PFs are a well-known method for computing upper bounds on the cost incurred in a single program~\cite{HoffmannAH11,HoffmannH10,HoffmannDW17,Carbonneaux0S15,0002AH12}. In this work, we also introduce {\em anti-potential functions (anti-PFs)}, a notion dual to potential functions that allow us to compute lower bounds on incurred cost.

Informally, a \emph{potential function (PF)} in a program is a function $\potential$ that assigns a real value to each program state (comprising of a location in the code together with the vector of variable values). It is required to satisfy two conditions, intuitively capturing sufficient resource to reach termination:
\begin{description}
	\item[Sufficiency preservation.] For any reachable state $\mathbf{c}$ in the program and any successor state $\mathbf{c}'$ of $\mathbf{c}$, we have
	\[ \potential(\mathbf{c})\geq \potential(\mathbf{c}') + \Delta_{\cost}(\mathbf{c},\mathbf{c}') \;,  \]
	where here we use $\Delta_{\cost}(\mathbf{c},\mathbf{c}')$ as informal notation for the cost incurred by progressing from $\mathbf{c}$ to $\mathbf{c}'$.

	\item[Sufficiency on termination.] The $\potential$ function is nonnegative upon program termination (i.e., $\potential(\mathbf{c}) \geq 0$ for a terminating state $\mathbf{c}$).
\end{description}
Intuitively, these properties impose that $\potential$ provides enough resources for a program run to progress to a successor state where the resources are used to ``pay'' for the incurred cost, and that the remaining amount of resources upon termination is nonnegative (i.e., sufficient to ``pay'' for the execution).

\emph{Anti-potential functions (anti-PFs)} are required to satisfy dual properties to PFs. An anti-PF $\antipotential$ assigns a real value to each program state and is required to satisfy two conditions, intuitively capturing \emph{in}sufficient resources to reach termination:
\begin{description}
	\item[Insufficiency preservation.] For any reachable state $\mathbf{c}$
	and any successor state $\mathbf{c}'$ of $\mathbf{c}$, we have
	\[ \antipotential(\mathbf{c})\leq \antipotential(\mathbf{c}') + \Delta_{\cost}(\mathbf{c},\mathbf{c}') \;.\]

	\item[Insufficiency on termination.] The $\antipotential$ function is nonpositive upon program termination (i.e., $\antipotential(\mathbf{c}) \leq 0$ for a terminating state $\mathbf{c}$).
\end{description}
Note that these properties for anti-PFs are indeed dual to those imposed by PFs: they require that $\antipotential$ {\em does not} provide enough resources for a program run to progress from any state to a successor state, and that the remaining amount of resources upon termination is {\em nonpositive} (i.e., insufficient to ``pay'' for the execution).

We formally define PFs and anti-PFs in Section~\ref{sec:potential} and then show that for any reachable state $\mathbf{c}$, the values $\potential(\mathbf{c})$ and $\antipotential(\mathbf{c})$ provide an upper and a lower bound on the cost incurred by any program run that starts in $\mathbf{c}$, respectively.

\begin{example}[Potentials and anti-potentials]\label{ex:runningpfs}
	Consider again our running example presented in Fig.~\ref{fig:running}. 
	Examples of a PF $\newsub{\potential}$ for the new version of \lstinline{join} and of an anti-PF $\oldsub{\antipotential}$ for the old version of \lstinline{join} are presented as annotations in Fig.~\ref{fig:running}.
	%
	To see that $\newsub{\potential}$ indeed defines a PF for the new version of \lstinline{join}, observe that a program run incurs cost $2$ in each inner loop iteration and otherwise does not incur cost, hence the value of the PF needs to decrease by at least $2$ with respect to the transition from $\loc_3$ to $\loc_2$, and to be non-increasing with respect to other transitions. The expression for $\newsub{\potential}$ can be seen to satisfy this property at any reachable state. Note that at any reachable state in the program, we have $\idxi \in [0,\lenA]$ and $\idxj \in [0,\lenB]$. Moreover, it is nonnegative upon termination. Hence, $\newsub{\potential}$ is a PF for the new version of \lstinline{join}.
	%
	One analogously verifies that $\oldsub{\antipotential}$ is nonpositive upon termination and that, upon executing each transition, the value of $\oldsub{\antipotential}$ increases at most by the incurred cost. Hence, $\oldsub{\antipotential}$ is an anti-PF in the old version of \lstinline{join}.
	
	Observe that in both cases, the expression that defines the PF and the anti-PF at each initial state evaluates to the exact cost usage of a program run that starts in that state. Hence, polynomial PFs and anti-PFs allow us to capture the exact cost usage in both versions of \lstinline{join}.
\end{example}

\JEDI{Explaining the idea of using PFs and anti-PFs for DiffCost. Demonstration on an example.}

\myparagraph{Using PFs and anti-PFs for differential cost analysis.} 
We now describe how PFs and anti-PFs in combination can be used to reason about the difference in cost between two programs.

Consider the DiffCost problem for two given programs and the set $\Theta_0$ of inputs. Suppose that we are able to compute a PF $\newsub{\potential}$ for the new program and an anti-PF $\oldsub{\antipotential}$ for the old program. Then, in Section~\ref{sec:potentialrelational}, we prove that for any input $\mathbf{c}\in\Theta_0$, the difference $\newsub{\potential}(\mathbf{c})-\oldsub{\antipotential}(\mathbf{c})$ is an {\em upper bound} on the difference in cost between the two programs on input $\mathbf{c}$. Hence, if a PF $\newsub{\potential}$ and an anti-PF $\oldsub{\antipotential}$ satisfy
\begin{equation}\label{eq:relcost}
\forall \mathbf{c}\in\Theta_0.\, \newsub{\potential}(\mathbf{c})-\oldsub{\antipotential}(\mathbf{c}) \leq t
\end{equation}
for value $t$,
then $t$ is a threshold for the DiffCost problem. This is the essence of our method for differential cost analysis. In particular, to compute a threshold $t$ for the DiffCost problem, our method computes the following three objects:
\begin{enumerate}
	\item a PF $\newsub{\potential}$ for the new program,
	\item an anti-PF $\oldsub{\antipotential}$ for the old program, and
	\item a value $t$ which together with $\newsub{\potential}$ and $\oldsub{\antipotential}$ satisfies the threshold bound (i.e., eq.~\eqref{eq:relcost} above).
\end{enumerate}
In Theorem~\ref{thm:potentialrelational} in Section~\ref{sec:potentialrelational}, we show that this approach to differential cost analysis is not only sound but also theoretically {\em complete}. In particular, we prove that whenever $t$ is a threshold for the DiffCost problem, there exists a PF $\newsub{\potential}$ and an anti-PF $\oldsub{\antipotential}$ for two given programs, which along with the threshold $t$ satisfy eq.~\eqref{eq:relcost}. We also show in
Theorem~\ref{thm:refutethreshold}, that PFs and anti-PFs can be used to prove that if a given value $t$ is {\em not} a threshold for the programs, then the difference $t$ in cost can be strictly exceeded on at least one input.

\begin{example}[Differential cost analysis with potentials and anti-potentials]\label{ex:runningrelcost}
	We now illustrate how this idea can be used to reason about the difference in cost between two programs in Fig.~\ref{fig:running}. Consider the PF $\newsub{\potential}$ and the anti-PF $\oldsub{\antipotential}$ defined in Example~\ref{ex:runningpfs}. We have that for each initial program state $\mathbf{c}$, the difference in cost of running the two procedure versions on input $\mathbf{c}$ is at most
	\begin{equation*}
	\newsub{\potential}(\mathbf{c}) - \oldsub{\antipotential}(\mathbf{c}) = \lenA \cdot \lenB \;.
	\end{equation*}
	Since initial variable values are constrained to satisfy $1 \leq \lenA \leq 100 \land 1 \leq \lenB \leq 100$ (as specified by the \lstinline{assume} statement in Fig.~\ref{fig:running}), we conclude that
	\begin{equation*}
	\newsub{\potential}(\mathbf{c}) - \oldsub{\antipotential}(\mathbf{c}) \leq 100\cdot 100 = 10000
	\end{equation*}
	holds for any initial state $\mathbf{c}$. Thus, $t=10000$ is a {\em threshold} for the DiffCost problem for the revision to the \lstinline{join} procedure shown in Fig.~\ref{fig:running}.
\end{example}

\JEDI{Explaining the key challenge behind turning this idea into a method for differential cost analysis.}

\myparagraph{Challenge: Deriving PFs and anti-PFs.}
The idea of using PFs and anti-PFs for differential cost analysis is not entirely surprising once we think of them as a means to obtain upper and lower bounds on incurred cost. However, the key challenge in designing a method for differential cost analysis based on this idea is effectively computing PFs and anti-PFs that would obtain tight threshold values for the DiffCost problem. A na\"ive approach would be to compute a PF for the new program and an anti-PF for the old program separately, and then to compute a threshold for them. However, such computations of the PF and the anti-PF would not take each other into account, which might lead to imprecision.

\myparagraph{Approach: Computing PFs and anti-PFs simultaneously with a threshold.}
The key conceptual novelty of our method that tackles this challenge is to {\em simultaneously} compute a threshold $t$ together with a PF $\newsub{\potential}$ and an anti-PF $\oldsub{\antipotential}$ that witness it---by employing a constraint solving-based approach.
\JEDI{Simultaneous computation can be reduced to constraint solving. Using Handelman's theorem then allows reducing the problem to an LP instance that can be efficiently solved and allows minimizing the threshold value.}
%
%
We now outline the key ideas behind our algorithm for computing a threshold $t$ together with a PF $\newsub{\potential}$ and an anti-PF $\oldsub{\antipotential}$ that witness it. Further details can be found in Section~\ref{sec:algorithm}.

For both programs, our method first fixes a polynomial template for a PF by fixing a symbolic polynomial expression over program variables for each location in the program. It also fixes a symbolic variable for the threshold $t$. Then, the defining properties of PFs and anti-PFs as well as the threshold condition in eq.~\eqref{eq:relcost} are all encoded as constraints over the symbolic template variables. This results in a system of constraints, and any solution to the system gives rise to a threshold $t$ as well as to a PF and an anti-PF that witness it.

However, a challenging aspect of constraint solving is that the resulting system of constraints involves both universal quantifiers (e.g., the $\forall\mathbf{c}\in\Theta_0$ in eq.~\eqref{eq:relcost}) and polynomial constraints over variables that are hard to solve. Hence, using quantifier elimination to solve these systems of constraints directly would be inefficient. In order to remove universal quantifiers as well as to avoid solving systems of polynomial constraints, our method uses Handelman's theorem~\cite{handelman1988representing}, a result on positive polynomials from algebraic geometry, to show that these constraints can be converted into {\em purely existentially quantified} and {\em linear} constraints over the symbolic template variables. This allows us to reduce the synthesis problem for the threshold and the witnessing PF and anti-PF to solving a system of linear constraints. These linear constraints can then be solved efficiently via an off-the-shelf linear programming (LP) solver. Furthermore, this encoding allows our method to optimize and compute as tight of a threshold as possible, by setting the LP optimization objective to minimize~$t$. In Section~\ref{sec:algorithm}, we show that our method can also be used to verify a given {\em symbolic polynomial bound} in terms of program variables on the difference in cost usage, which may be achieved by dropping the minimization objective and replacing the concrete threshold $t$ with the polynomial bound of interest.

\JEDI{Caveat of our approach: precision guarantees on bounds for single programs.}

\myparagraph{Additional Benefit: Precision guarantees on bounds for a single program.}
While the motivation for this work is to address the differential cost analysis problem, an additional consequence of our approach is that it suggests an approach to compute upper and lower bounds on cost in a single program with {\em precision guarantees} on the computed bounds.
%
In particular, we show in Section~\ref{sec:tight} that if we simultaneously compute a PF and an anti-PF for a single program and we regard the threshold $t$ as the maximal difference between the two bounds, then it provides a bound on the {\em precision} of the computed bounds.

%% file: prelims.tex
\section{Preliminaries}\label{sec:prelims}

In this work, we consider imperative arithmetic programs with polynomial integer arithmetic. These allow standard programming constructs such as (polynomial) variable assignments, conditional branching and loops. In addition, we allow constructs for {\em non-deterministic} variable assignments.
Cost in programs is modeled by a special program variable $\cost$, which is initialized to $0$ and modified whenever cost is incurred in the program. 



\myparagraph{Predicates.} Given a finite set of (integer) variables $\vars$, a variable valuation is a vector $\mathbf{x}\in \mathbb{Z}^{|\vars|}$. A {\em predicate} over $\vars$ is a set of variable valuations (i.e., a subset of $\mathbb{Z}^{|\vars|}$). A predicate is said to be a {\em polynomial assertion} if it is a conjunction of finitely many polynomial inequalities over variables in $\vars$. If $\mathbf{x}\in\mathbb{Z}^{|\vars|}$ and $\phi$ is a predicate over $\vars$ given by a boolean formula, we write $\mathbf{x}\models\phi$ to denote that the formula $\phi$ is satisfied by substituting values of the components of $\mathbf{x}$ for the corresponding variables in $\phi$.

\myparagraph{Model for programs.} We model programs via transition systems. A {\em transition system}~\cite{ColonSS03} is a tuple $\transys=(\locs,\vars,\trans,\locinit,\initvars)$, where:
\begin{compactitem}
	\item $\locs$ is a finite set of {\em program locations}.
	\item $\vars$ is a finite set of {\em program variables}. We assume that each program has a distinguished variable $\cost$.
	\item $\trans$ is a finite set of {\em transitions}, which are tuples of the form $\tau=(\loc,\loc',\Guard^{\tau},\Update^{\tau})$. Here, $\loc$ is the {\em source} and $\loc'$ is the {\em target location} of $\tau$. $\Guard^{\tau}$ is the {\em guard} of $\tau$, and we assume that it is given by a polynomial assertion over variables $\vars$. Finally, $\Update^{\tau}$ is the {\em update} of $\tau$, which to each variable $v\in \vars$ assigns either a polynomial expression over $\vars$ with $\Update^{\tau}(v)=v$ if $\tau$ does not update $v$, or the set $\mathbb{Z}$ of integer numbers in case of a non-deterministic variable update.
	\item $\locinit$ denotes the {\em initial program location}.
	\item $\initvars$ denotes the set of {\em initial variable valuations}. We assume that $\initvars$ is a polynomial assertion and that for each initial variable valuation, we have $\cost=0$.
\end{compactitem}

We assume the existence of a special {\em terminal location} $\locterm$, which represents the final line of the program code. It has a single outgoing transition $(\locterm,\locterm,\text{true},\Update)$ with $\Update(v)=v$ for each $v\in\vars$. Furthermore, we assume that each location $\loc\in\locs$ has at least one outgoing transition and that it is always possible to execute at least one transition. This is done without loss of generality and may be enforced by introducing a dummy transition from $\loc$ to $\locterm$.

Translation of numerical integer programs into transition systems is standard, so we omit the details. However, for completeness in presentation, we do give the transition systems that model the procedures in Fig.~\ref{fig:running} in Appendix~\ref{app:transitionsystem}.

A {\em state} of a transition system is an ordered pair $(\loc,\mathbf{x})$ where $\loc\in\locs$ and $\mathbf{x}\in \mathbb{Z}^{|\vars|}$. A state is said to be {\em initial} if it is of the form $(\locinit,\varinit)$ with $\varinit\in\initvars$. A state is said to be {\em terminal} if it is of the form $(\locterm,\mathbf{x})$. A state $(\loc',\mathbf{x}')$ is a {\em successor} of a state $(\loc,\mathbf{x})$ if there exists a transition $\tau=(\loc,\loc',\Guard^{\tau},\Update^{\tau})$ with $\mathbf{x}\models\Guard^{\tau}$ and $\Update^{\tau}(v)(\mathbf{x}) = \mathbf{x}'[v]$ for each $v\in V$ whose update is deterministic. Given a state $\mathbf{c}$, a {\em finite path from $\mathbf{c}$} in $\transys$ is a finite sequence of states $\mathbf{c}_0=\mathbf{c},\mathbf{c}_1,\dots,\mathbf{c}_k$, where for each $0\leq i<k$ we have that $\mathbf{c}_{i+1}$ is a successor of $\mathbf{c}$. A state $\mathbf{c}'$ is {\em reachable from $\mathbf{c}$} if there exists a finite path from $\mathbf{c}$ that ends in $\mathbf{c}'$. A {\em run (or execution) from $\mathbf{c}$} is an infinite sequence of program states where each finite prefix is a finite path from $\mathbf{c}$. When we omit specifying a state $\mathbf{c}$, we refer to a finite path, run or reachability from some initial state.

For a state $\mathbf{c}$, we define $\mathit{Run}(\mathbf{c})$ to be the set of all runs in $\transys$ that start in $\mathbf{c}$. We denote by $\mathit{Run}_{\transys}$ to be the set of all runs in $\transys$ that start in some initial state of $\transys$.

\myparagraph{Cost of a run and the termination assumption.} In order for the notion of the cost of a run to be well-defined, in the rest of this work we assume that all our transition systems (and thus programs) are terminating. Given a transition system $\transys$ and a state $\mathbf{c}$, we say that a run from $\mathbf{c}$ is {\em terminating} if it has a finite prefix with the last state being terminal. We then say that the transition system $\transys$ is {\em terminating} if every run in $\transys$ from some initial state is terminating.

Given a run $\rho$ from some state $(\loc,\mathbf{x})$, let $(\locterm,\mathbf{x}_{\textit{out}})$ be the terminal state reached in $\rho$. Then the {\em cost} of $\rho$, denoted by $\Cost_{\transys}(\rho)$, equals the difference of the terminal and the initial value of the variable $\cost$ along $\rho$ (i.e., $\mathbf{x}_{\textit{out}}[\cost] - \mathbf{x}[\cost]$). We define the {\em maximal cost} of $(\loc,\mathbf{x})$ to be the maximal cost incurred by any run in $\transys$ from $(\loc,\mathbf{x})$, that is,
\[ \CostSup_{\transys}(\loc,\mathbf{x})=\sup\Big \{\Cost_{\transys}(\rho)\mid \rho\in \Run(\loc,\mathbf{x})\Big\} \;. \]
Similarly, we define the {\em minimal cost} of $(\loc,\mathbf{x})$ to be
\[ \CostInf_{\transys}(\loc,\mathbf{x})=\inf\Big\{\Cost_{\transys}(\rho)\mid \rho\in \Run(\loc,\mathbf{x})\Big\} \;. \]
Note that these two values might differ due to the possible existence of non-determinism in the program. However, we always have $\CostInf_{\transys}(\loc,\mathbf{x})\leq \CostSup_{\transys}(\loc,\mathbf{x})$.

The termination assumption is essential for the cost of a run to be well-defined. Indeed, for a non-terminating run one can naively try to define its cost by taking the limit of costs of its finite prefixes. However, this limit does not need to necessarily exist as we allow both positive and negative costs. Allowing costs of arbitrary sign is necessary to model some of the most important use cases of cost analysis. For instance, in order to obtain tight bounds on memory usage in programs, we need to take into account that memory can be released and returned to the program which is modeled by incurring negative cost.


\myparagraph{The differential cost analysis problem.} We now formally define the differential cost analysis problem that we consider in this work. Given two programs, our goal is to compare their cost usage and to compute a bound on the maximal difference in incurred cost between the two programs given the same input.

Formally, let $\newsub{\transys} = (\newsub{\locs},\vars,\newsub{\trans},\newsub{\locinit},\initvars)$ and $\oldsub{\transys}=(\oldsub{\locs},\vars,\oldsub{\trans},\oldsub{\locinit},\initvars)$ be two transition systems that share the same finite set of variables $\vars$ and the set $\initvars$ of initial variable valuations. The {\em differential cost analysis (DiffCost) problem} asks to compute a {\em threshold} $t\in\mathbb{Z}$ on the difference in cost usage that cannot be exceeded, that is, $t\in\mathbb{Z}$ for which the following logical formula is true:
\[ \forall \mathbf{x}\in \initvars.\, \CostSup_{\newsub{\transys}}(\newsub{\loc_0},\mathbf{x}) - \CostInf_{\oldsub{\transys}}(\oldsub{\loc_0},\mathbf{x}) \leq t. \]

%% file: potentialfunctions.tex
\section{Potential and Anti-potential Functions for Differential Cost Analysis}\label{sec:potentialsection}

Potential functions (PFs) from amortized analysis~\cite{tarjan1985amortized} are a classical method for computing upper bounds on the cost incurred in a single program. In this work we show that PFs and their dual for computing lower bounds on incurred cost, which we call {\em anti-potential functions (anti-PFs)}, can also be used to reason about differential cost analysis.

\subsection{Potential and Anti-potential Functions}\label{sec:potential}

Let $\transys$ be a transition system. A {\em potential function (PF)} in $\transys$ is a map $\potential$ that assigns a real value to each state in $\transys$, and which satisfies the following two properties:
\begin{description}
	\item[Sufficiency preservation.] For any reachable state $(\loc,\mathbf{x})$ in $\transys$ and any successor state $(\loc ',\mathbf{x}')$ of $(\loc,\mathbf{x})$, we have
	\[ \potential(\loc,\mathbf{x})\geq \potential(\loc',\mathbf{x}') + \mathbf{x}'[\cost] - \mathbf{x}[\cost].  \]
	We use $\mathbf{x}[\cost]$ and $\mathbf{x}'[\cost]$ to denote the values of the variable $\cost$ defined by valuations $\mathbf{x}$ and $\mathbf{x}'$.
	
	\item[Sufficiency on termination.] For any reachable terminal state $(\locterm,\mathbf{x})$ in $\transys$, we have $\potential(\locterm,\mathbf{x})\geq 0$.
\end{description}

One can define a notion dual to PFs in order to compute lower bounds on the cost usage of a given program, provided that the program is terminating. An {\em anti-potential function (anti-PF)} in $\transys$ is a map $\antipotential$, which to each state in $\transys$, assigns a real value, and which satisfies the following two properties:
\begin{description}
	\item[Insufficiency preservation.] For any reachable state $(\loc,\mathbf{x})$ in $\transys$ and any successor state $(\loc ',\mathbf{x}')$ of $(\loc,\mathbf{x})$, we have
	\[ \antipotential(\loc,\mathbf{x})\leq \antipotential(\loc',\mathbf{x}') + \mathbf{x}'[\cost] - \mathbf{x}[\cost].  \]
	\item[Insufficiency on termination.] For any reachable terminal state $(\locterm,\mathbf{x})$ in $\transys$, we have $\antipotential(\locterm,\mathbf{x})\leq 0$.
\end{description}

The following theorem shows that, given a reachable state $(\loc,\mathbf{x})$ in $\transys$, PFs and anti-PFs evaluate to upper bounds on the maximal cost and lower bounds on the minimal cost of a run starting $(\loc,\mathbf{x})$, respectively. Observe that, in order to prove these inequalities, it suffices to prove that for any run $\rho$ starting in $(\loc,\mathbf{x})$, the values $\potential(\loc,\mathbf{x})$ and $\antipotential(\loc,\mathbf{x})$ are respectively an upper and a lower bound on the cost of $\rho$. We prove this by induction on the length of $\rho$, and the proof can be found in Appendix~\ref{app:proofs}.

\begin{theorem}\label{thm:potential}
	Let $\transys$ be a transition system that is terminating. If $\potential$ is a PF in $\transys$, then for any reachable state $(\loc,\mathbf{x})$ in $\transys$, we have
	\[ \potential(\loc,\mathbf{x}) \geq \CostSup_{\transys}(\loc,\mathbf{x}). \]
	If $\antipotential$ is an anti-PF in $\transys$, then for any reachable state $(\loc,\mathbf{x})$ in $\transys$ we have
	\[ \antipotential(\loc,\mathbf{x}) \leq \CostInf_{\transys}(\loc,\mathbf{x}). \]
\end{theorem}

Recall, Example~\ref{ex:runningpfs} in Section~\ref{sec:overview} shows a pair of a PF and an anti-PF in two versions of the procedure \lstinline{join} in Fig.~\ref{fig:running}.

We note that the termination assumption is necessary for Theorem~\ref{thm:potential} to hold, as the claim for anti-PFs may be violated even if all incurred costs are of the same sign so that the cost of a run in a non-terminating program is well-defined. An example program demonstrating this necessity is provided in Appendix~\ref{app:nonterm}.


\subsection{Application to Differential Cost Analysis}\label{sec:potentialrelational}

We now proceed to show how PFs and anti-PFs can be used to reason about differential cost analysis. Let $\newsub{\transys} = (\newsub{\locs},\vars,\newsub{\trans},\newsub{\locinit},\initvars)$ and $\oldsub{\transys}=(\oldsub{\locs},\vars,\oldsub{\trans},\oldsub{\locinit},\initvars)$ be two terminating transition systems that share the finite set of variables $\vars$ and the set $\initvars$ of initial variable valuations. The following theorem shows that PFs and anti-PFs are {\em sound} for computing the threshold on the maximal difference in cost usage for the DiffCost problem that we defined in Section~\ref{sec:prelims}. It also shows that they are theoretically {\em complete}, in the sense that any valid threshold for the DiffCost problem admits a pair of a PF in $\newsub{\transys}$ and an anti-PF in $\oldsub{\transys}$ that witness it.

The proof of soundness follows from Theorem~\ref{thm:potential}, and the proof of completeness follows by observing that the maximal and minimal costs for each state satisfy the defining properties of PFs and anti-PFs. We defer the proof to Appendix~\ref{app:proofs}.

\begin{theorem}[PFs and anti-PFs for DiffCost]\label{thm:potentialrelational}
	Let $\newsub{\transys}$ and $\oldsub{\transys}$ be two terminating transition systems. Suppose that $\newsub{\potential}$ is a PF in $\newsub{\transys}$ and that $\oldsub{\antipotential}$ is an anti-PF in $\oldsub{\transys}$. Then, for each initial variable valuation $\mathbf{x}\in\initvars$, we have that
	\begin{equation*}
	\begin{split}
	&\CostSup_{\newsub{\transys}}(\newsub{\loc_0},\mathbf{x}) - \CostInf_{\oldsub{\transys}}(\oldsub{\loc_0},\mathbf{x}) \\
	&\leq \newsub{\potential}(\newsub{\locinit},\mathbf{x}) - \oldsub{\antipotential}(\oldsub{\locinit},\mathbf{x}).
	\end{split}
	\end{equation*}
	In particular, if $t$ satisfies $\newsub{\potential}(\newsub{\locinit},\mathbf{x}) - 
	\oldsub{\antipotential}(\oldsub{\locinit},\mathbf{x}) \leq t$ for each $\mathbf{x}\in\initvars$, then $t$ is a threshold for the DiffCost problem.
	
	Conversely, if $t\in\mathbb{Z}$ is a threshold for the DiffCost problem, then there exist a PF $\newsub{\potential}$ in $\newsub{\transys}$ and an anti-PF $\oldsub{\antipotential}$ in $\oldsub{\transys}$ such that $\newsub{\potential}(\newsub{\locinit},\mathbf{x}) - 
	\oldsub{\antipotential}(\oldsub{\locinit},\mathbf{x}) \leq t$ holds for each $\mathbf{x}\in\initvars$.
\end{theorem}

Hence, in order to compute a threshold $t$ for the DiffCost problem, it suffices to compute the following three objects:
\begin{enumerate}
	\item a PF $\newsub{\potential}$ for $\newsub{\transys}$,
	\item a anti-PF $\oldsub{\antipotential}$ for $\oldsub{\transys}$, and
	\item an integer $t\in\mathbb{Z}$ which together with $\newsub{\potential}$ and $\oldsub{\antipotential}$ satisfies $\newsub{\potential}(\newsub{\locinit},\mathbf{x}) - \oldsub{\antipotential}(\oldsub{\locinit},\mathbf{x}) \leq t$ for each initial variable valuation $\mathbf{x}\in\initvars$.
\end{enumerate}
We already illustrated this idea in Example~\ref{ex:runningrelcost}, Section~\ref{sec:overview}, on our running example in Fig.~\ref{fig:running}.

\subsection{Refuting a Threshold via PFs and Anti-PFs}

We conclude this section by showing that PFs and anti-PFs can also be used to prove that some threshold in differential cost can be {\em strictly exceeded}. We defer the proof to Appendix~\ref{app:proofs}. While the approach for refuting a threshold value is sound for general programs with non-determinism, it is complete only for deterministic programs. 

\begin{theorem}[Refuting a threshold]\label{thm:refutethreshold}
	Let $\newsub{\transys}$ and $\oldsub{\transys}$ be two terminating transition systems. Suppose that $\newsub{\antipotential}$ is an anti-PF in $\newsub{\transys}$, and that $\oldsub{\potential}$ is a PF in $\oldsub{\transys}$. Then, for each initial variable valuation $\mathbf{x}\in\initvars$, we have that
	\begin{equation*}
	\begin{split}
	&\CostInf_{\newsub{\transys}}(\newsub{\loc_0},\mathbf{x}) - \CostSup_{\oldsub{\transys}}(\oldsub{\loc_0},\mathbf{x}) \\
	&\geq \newsub{\antipotential}(\newsub{\locinit},\mathbf{x}) - \oldsub{\potential}(\oldsub{\locinit},\mathbf{x}).
	\end{split}
	\end{equation*}
	In particular, if $t\in\mathbb{Z}$ satisfies $\newsub{\antipotential}(\newsub{\locinit},\mathbf{x}) - \oldsub{\potential}(\oldsub{\locinit},\mathbf{x}) > t$ for some $\mathbf{x}\in\initvars$, then $t$ is {\em not} a threshold for the DiffCost problem.
	
	Conversely, if $t$ is not a threshold for the DiffCost problem and if $\newsub{\transys}$ and $\oldsub{\transys}$ are induced by deterministic programs, then there exist an anti-PF $\newsub{\antipotential}$ in $\newsub{\transys}$ and a PF $\oldsub{\potential}$ in $\oldsub{\transys}$ such that $\newsub{\antipotential}(\newsub{\locinit},\mathbf{x}) - \oldsub{\potential}(\oldsub{\locinit},\mathbf{x}) > t$ for at least one $\mathbf{x}\in\initvars$.
\end{theorem}

\begin{example}
	To illustrate how PFs and anti-PFs can be used to prove that some threshold in cost difference can be exceeded, consider again our running example in Fig.~\ref{fig:running}, and the PF $\newsub{\potential}$ and the anti-PF $\oldsub{\antipotential}$ that were defined in Example~\ref{ex:runningpfs}. By analogous reasoning as in Example~\ref{ex:runningpfs}, one may easily check that $\newsub{\antipotential}=\newsub{\potential}$ is also an anti-PF in the new version of \lstinline{join} and that $\oldsub{\potential}=\oldsub{\antipotential}$ is a PF in the old version of \lstinline{join}. On the other hand, from the \lstinline{assume} statements in Fig.~\ref{fig:running} we see that the initial variable valuations in both programs are given by the assertion $\Theta_0=1\leq \lenA\leq 100 \land 1\leq \lenB\leq 100$. Hence, for any initial variable valuation $\mathbf{x}\in\Theta_0$, we have
	\begin{equation*}
	\begin{split}
	&\CostInf_{\newsub{\transys}}(\newsub{\loc_0},\mathbf{x}) - \CostSup_{\oldsub{\transys}}(\oldsub{\loc_0},\mathbf{x}) \\
	&\geq \newsub{\antipotential}(\newsub{\locinit},\mathbf{x}) - \oldsub{\potential}(\oldsub{\locinit},\mathbf{x}) = \lenA \cdot \lenB.
	\end{split}
	\end{equation*}
	Thus, as $\lenA \cdot \lenB = 10000$ for any initial variable valuation in $\Theta_0$ with $\lenA=100$ and $\lenB=100$, it follows that, $t=9999$ is not a threshold for the DiffCost problem for our running example in Fig.~\ref{fig:running}.
\end{example}

We elaborate on the reason why PFs and anti-PFs are complete for refuting threshold values only in deterministic programs. Suppose that $\newsub{\antipotential}(\newsub{\locinit},\mathbf{x}) - \oldsub{\potential}(\oldsub{\locinit},\mathbf{x}) > t$ holds for some $\mathbf{x}\in\initvars$. Then, for {\em every run} $\newsub{\rho}$ starting in $(\newsub{\locinit},\mathbf{x})$ in $\newsub{\transys}$ and {\em every run} $\oldsub{\rho}$ starting in $(\oldsub{\locinit},\mathbf{x})$ in $\oldsub{\transys}$, the cost of $\newsub{\rho}$ exceeds the cost of $\oldsub{\rho}$ by an amount that is strictly greater than $t$. However, for $t$ not to be a threshold value, it would suffice that this happens for a single pair of such runs. PFs and anti-PFs cannot be used to witness this weaker condition. In the case of deterministic programs, however, there is only a single run $\newsub{\rho}$ starting in $(\newsub{\locinit},\mathbf{x})$ and a single run $\oldsub{\rho}$ starting in $(\oldsub{\locinit},\mathbf{x})$, hence PFs and anti-PFs are complete for refuting threshold values in deterministic programs.

%% file: algorithm.tex
\section{Simultaneous Potentials and Anti-potentials Algorithm}\label{sec:algorithm}

We now present our algorithm for differential cost analysis. Our algorithm is based on the idea that was presented in Section~\ref{sec:potentialrelational}, and it simultaneously computes a polynomial PF for the new program, a polynomial anti-PF for the old program, as well as a threshold value. The algorithm runs in polynomial time, and reduces the computation of the PF, the anti-PF and the threshold value to a linear programming (LP) instance by employing a constraint solving-based approach. 

In what follows, let $\newsub{\transys} = (\newsub{\locs},\vars,\newsub{\trans},\newsub{\locinit},\initvars)$ and $\oldsub{\transys}=(\oldsub{\locs},\vars,\oldsub{\trans},\oldsub{\locinit},\initvars)$ be two terminating transition systems that share the finite set of variables $\vars$ and the set $\initvars$ of initial variable valuations. 

\smallskip\noindent{\em Algorithm assumptions.} Our algorithm has two constant natural number parameters $d, K\in\mathbb{N}$, where $d$ is the maximal degree of polynomials that it considers and $K$ is a parameter whose meaning we will explain shortly. It also assumes the following:
\begin{asparaenum}
	\item {\em Affine invariants.} Recall that the defining properties of PFs and anti-PFs impose conditions on their values at {\em reachable} states in programs. In order to compute PFs and anti-PFs, our algorithm assumes that it is provided with {\em invariants} $\newsub{I}$ for $\newsub{\transys}$ and $\oldsub{I}$ for $\oldsub{\transys}$. An invariant is an over-approximation of the set of all reachable program states. Formally, an invariant in a program is a map $I$ that to each program location $\loc$ assigns a predicate $I(\loc)$, such that for any reachable state $(\loc,\mathbf{x})$ in the program we have $\mathbf{x}\models I(\loc)$. 
	We assume that the pre-computed invariants are {\em affine}, meaning that each $I(\loc)$ is given by a conjunction of finitely many affine inequalities over program variables. Affine invariant generation is a well-studied problem in program analysis, with several efficient methods and tools~\cite{FeautrierG10,SankaranarayananSM04}. 
	
	\item Each transition guard is assumed to be given in terms of affine inequalities. This assumption is made without loss of generality, as any non-affine expression may be replaced by a dummy variable to which the value of this expression is previously assigned.
	
	\item $\Theta_0$ is assumed to be a conjunction of affine inequalities.
\end{asparaenum}

\smallskip\noindent{\em Constraint solving-based approach.} In order to compute a threshold value for the DiffCost problem, our algorithm employs a constraint solving-based approach to simultaneously search for a polynomial PF $\newsub{\potential}$ in $\newsub{\transys}$, a polynomial anti-PF $\oldsub{\antipotential}$ in $\oldsub{\transys}$, and a threshold value $t$; it proceeds in $4$ steps. First, the algorithm fixes symbolic polynomial templates for $\newsub{\potential}$ and $\oldsub{\antipotential}$, as well as a symbolic template variable for $t$. Second,  the algorithm collects the defining properties of PFs, anti-PFs and the differential cost constraint. Third, the collected constraints are soundly converted into a system of purely existentially quantified linear constraints. Fourth, the resulting system of linear constraints is efficiently solved by an off-the-shelf LP solver. The threshold value is in addition minimized, in order to compute a bound as tight as possible on relative cost. In what follows, we describe each of these $4$ steps in more detail.

\smallskip\noindent{\em Step 1: Symbolic templates.} The algorithm fixes a symbolic polynomial template of degree at most $d$ for $\newsub{\potential}$ by introducing symbolic polynomial $\newsub{\potential}(\loc)$ of degree at most $d$ for each location $\loc\in\newsub{\locs}$. This is done as follows. Let $\text{Mono}_d(\vars)$ be the set of all monomials of degree at most $d$ over the variable set $\vars$. Then, the symbolic template for $\newsub{\potential}(\loc)$ is a symbolic polynomial expression $\sum_{f\in \text{Mono}_d(\vars)}u^{\loc}_f\cdot f$, where $u^{\loc}_f$ is a real-valued symbolic template variable for each $f\in\text{Mono}_d(\vars)$.

Similarly, the algorithm fixes a symbolic polynomial template of degree at most $d$ for $\oldsub{\antipotential}$. Finally, it fixes a real-valued symbolic template variable $t$ for the threshold value.

\smallskip\noindent{\em Step 2: Constraint collection.} The algorithm now collects all the defining constraints for $\newsub{\potential}$ to be a PF in $\newsub{\transys}$, for $\oldsub{\antipotential}$ to be an anti-PF in $\oldsub{\transys}$ and for $t$ to be a threshold value:
\begin{enumerate}
	\item {\em PF constraints.} Recall that a PF needs to satisfy the \emph{sufficiency} preservation condition at all reachable states and the sufficiency on termination condition at all reachable terminal states. To capture that a state is reachable, our algorithm collects constraints which impose the defining conditions at all states contained in the invariant $\newsub{I}$. In particular, our algorithm collects the following constraints:
	\begin{compactitem}
		\item For each transition $\tau = (\loc,\loc',G^\tau,\Update^{\tau})\in\,\newsub{\trans}$,
		\begin{equation*}
		\begin{split}
		\mathbf{x}\models \newsub{I}(\loc)\cap G^\tau \Rightarrow &\,\newsub{\potential}(\loc,\mathbf{x}) \geq
		\newsub{\potential}(\loc',\Update^\tau(\mathbf{x})) \\
		&+  \Update^\tau(\cost)(\mathbf{x}) - \mathbf{x}[\cost].
		\end{split}
		\end{equation*}
		Here, $\newsub{\potential}(\loc',\Update^\tau(\mathbf{x}))$ is a notation for the expression obtained by taking the template polynomial for $\newsub{\potential}$ at $\loc'$ and substituting for each variable $v\in\vars$ either the polynomial update $\Update^{\tau}(v)(\mathbf{x})$, or a fresh variable in the case of a non-deterministic update.
		\item $\mathbf{x}\models \newsub{I}(\locterm) \Rightarrow \newsub{\potential}(\locterm,\mathbf{x}) \geq 0$.
	\end{compactitem}
	\item {\em Anti-PF constraints.} Similarly, an anti-PF needs to satisfy the \emph{insufficiency} preservation and the insufficiency on termination conditions, so our algorithm collects the following constraints:
	\begin{compactitem}
		\item For each transition $\tau = (\loc,\loc',G^\tau,\Update^{\tau})\in\,\oldsub{\trans}$,
		\begin{equation*}
		\begin{split}
		\mathbf{x}\models \oldsub{I}(\loc)\cap G^\tau \Rightarrow &\,\oldsub{\antipotential}(\loc,\mathbf{x}) \leq
		\oldsub{\antipotential}(\loc',Up^\tau(\mathbf{x})) \\
		&+  \Update^\tau(\cost)(\mathbf{x}) - \mathbf{x}[\cost].
		\end{split}
		\end{equation*}
		where $\oldsub{\antipotential}(\loc',\Update^\tau(\mathbf{x}))$ is analogous to $\newsub{\potential}(\loc',\Update^\tau(\mathbf{x}))$.
		\item $\mathbf{x}\models \oldsub{I}(\locterm) \Rightarrow \oldsub{\antipotential}(\locterm,\mathbf{x}) \leq 0$.
	\end{compactitem}
	\item {\em Differential cost constraint.} Our algorithm then collects the differential cost constraint:
	\[ \mathbf{x} \models \initvars \Rightarrow \newsub{\potential}(\newsub{\locinit},\mathbf{x}) - \oldsub{\antipotential}(\oldsub{\locinit},\mathbf{x}) \leq t. \]
\end{enumerate}
Observe that for each collected constraint, the expressions on the right-hand--side are linear in symbolic template variables for $\newsub{\potential}$, $\oldsub{\antipotential}$ and $t$, and the expressions on the left-hand--side contain no symbolic template variables.

\smallskip\noindent{\em Step 3: Conversion to a linear program.} The algorithm now converts each collected constraint into a system of purely existentially quantified linear constraints. To do this, we observe that each collected constraint has the following form:
\begin{equation}\label{eq:constraint}
\text{aff}_1(\mathbf{x})\geq 0 \land \dots \land \text{aff}_k(\mathbf{x})\geq 0 \Rightarrow \text{poly}(\mathbf{x})\geq 0,
\end{equation}
for some $k\geq 0$, where each $\text{aff}_i$ is an affine expression and $\text{poly}$ is a polynomial expression over program variables. This is because the left-hand-side of each constraint collected in Step~2 depends either on program invariants, transition guards or the initial variable valuation set $\initvars$, all of which are assumed to be affine. 

The algorithm converts the constraint in eq.~\eqref{eq:constraint} into a system of linear constraints by requiring $\text{poly}$ to be equal to a non-negative linear combination of finitely many products of affine expressions in $\text{Aff} = \{\text{aff}_1,\dots,\text{aff}_k\}$. To formalize this, we define $\text{Prod}_K(\text{Aff})$ to be the set of products of at most $K$ affine expressions in $\text{Aff}$, i.e.
\[ \text{Prod}_K(\text{Aff}) = \Big\{ \prod_{i=1}^t a_i \,\Big|\, t\in\mathbb{N}_0,\, t\leq d,\, a_1,\dots,a_t\in\text{Aff} \Big\}. \]
Recall that $K$ is one of the algorithm's two natural number parameters, and we introduce it in order to bound the maximal number of affine expressions that may appear in each product. One can see that any $g\in \text{Prod}_K(\text{Aff})$ satisfies $g(\mathbf{x})\geq 0$ for any valuation $\mathbf{x}$ in which $\text{aff}_i(\mathbf{x})\geq 0$ for each $\text{aff}_i\in\text{Aff}$. Hence, one can soundly translate the constraint in eq.~\eqref{eq:constraint} into another constraint that encodes that $\text{poly}$ can be written as a non-negative linear combination of finitely many products of affine expressions in $\text{Aff}$:
\begin{equation}\label{eq:handelmanaffine}
\text{poly}(\mathbf{x}) = \sum_{g \in \text{Prod}_K(\text{Aff})} c_g \cdot g(\mathbf{x}),
\end{equation}
where each $c_g\geq 0$.

To encode eq.~\eqref{eq:handelmanaffine} as a system of linear constraints, our algorithm introduces a fresh symbolic variable $c_g$ and a constraint $c_g\geq 0$ for each $g \in \text{Prod}_K(\text{Aff})$. Then, for each monomial over the variable set $V$ it equalizes the coefficients of the monomial on two sides of eq.~\eqref{eq:handelmanaffine} to produce a linear constraint over the symbolic template variables for $\newsub{\potential}$, $\oldsub{\antipotential}$ and $t$, as well as the symbolic variables $c_g$. Collecting all the produced constraints results in a sound translation of the constraint in eq.~\eqref{eq:constraint} into a system of purely existentially quantified linear constraints over symbolic variables.


While the translation of eq.~\eqref{eq:constraint} into eq.~\eqref{eq:handelmanaffine} is sound, it is not necessarily complete. However, Handelman's theorem implies that, if we work over real arithmetic and if the set $\langle\text{Aff}\,\rangle = \{\mathbf{x}\in\mathbb{R}^{\vars} \mid \text{aff}_i(\mathbf{x})\geq 0 \text{ for each } \text{aff}_i\in\text{Aff} \}$ is topologically compact (i.e., closed and bounded), satisfiability of eq.~\eqref{eq:handelmanaffine} is also a necessary condition for $\text{poly}$ to be strictly positive over~$\langle\text{Aff}\,\rangle$.

\smallskip\noindent\textbf{Handelman's Theorem}~\cite{handelman1988representing}.
\emph{
	Let $V$ be a finite set of real variables and $\text{Aff} = \{\text{aff}_1,\dots,\text{aff}_k\}$ be a set of finitely many affine expressions over $V$ (degree $1$ polynomials). Let $f\in\mathbb{R}[V]$ be a polynomial and suppose that $f(\mathbf{x})>0$ for all $\mathbf{x}\in\langle\text{Aff}\,\rangle = \{\mathbf{x}\in\mathbb{R}^{V} \mid \text{aff}_i(\mathbf{x})\geq 0 \text{ for each } \text{aff}_i\in\text{Aff} \}$. 
	If $\langle\text{Aff}\,\rangle$ is a compact set, then there exist $d\in \mathbb{N}_0$ and $s\in \mathbb{N}_0$ such that
	\begin{equation*}\label{eq:handelman}
	f = \sum_{i=1}^s c_i\cdot g_i
	\end{equation*}
	for some $c_1,\dots,c_s\geq 0$ and $g_1,\dots,g_s\in \text{Prod}_K(\Gamma)$.
}
\smallskip

Hence, if $\langle\text{Aff}\,\rangle$ is a compact set, the translation of eq.~\eqref{eq:constraint} into eq.~\eqref{eq:handelmanaffine} is also a complete method to impose a slightly different constraint than the one in eq.~\eqref{eq:constraint}, with the strict inequality $>$ instead of $\geq$ on the right-hand--side and the inequalities on the left-hand--side being imposed for all real-valued valuations $\mathbf{x}$ (and not just integer-valued). Note that $\langle\text{Aff}\,\rangle$ being compact is an assumption that is satisfied whenever all variable values in programs are bounded, including the variable $\cost$ that tracks the total resource usage. 
Indeed, if we know variable value bounds then we may add these bounds to the invariant at each program location as well as to $\Theta_0$. Since the set $\text{Aff}$ induced by any constraint collected in Step~2 contains either inequalities that define an invariant at some program location or inequalities that define $\Theta_0$, this modification would result in $\text{Aff}$ being compact.

\smallskip\noindent{\em Step 4: Synthesis via constraint solving.} Denote by $\Phi$ the set of all linear constraints produced in Step~3. The algorithm employs an off-the-shelf LP solver to solve
\begin{equation*}
\begin{split}
&\text{minimize } t\\
&\text{subject to } \Phi \;.
\end{split}
\end{equation*}
The algorithm then outputs the computed $t$ if the LP solver finds a solution, or ``Unknown'' otherwise.

\smallskip The following theorem establishes soundness of our algorithm and that it runs in polynomial time. The proof is provided in Appendix~\ref{app:proofs}.

\begin{theorem}[Soundness]
	If the algorithm outputs a value $t$, then $t$ is a threshold for the DiffCost problem. Furthermore, the algorithm runs in polynomial time for the fixed values of parameters $d$ and $K$.
\end{theorem}

	Our method computes polynomial PFs and anti-PFs and hence the computed cost bounds are also polynomials over program variables. However, it is known that computing tight cost bounds in programs with more complex control-flow may require disjunctive expressions involving piecewise-linear operators such as $\max$ or $\min$~\cite{GulwaniZ10,Carbonneaux0S15,SinnZV17}. Hence, for such programs, it might not be possible to compute polynomial cost bounds that provide desired precision guarantees. 
	
	Computing disjunctive cost bounds has been the focus of several works on single program cost analysis. In particular, the work of \citet{Carbonneaux0S15} also employs a constraint solving-based approach to compute {\em piecewise linear upper bounds} on cost usage in programs. An interesting direction for future work would be to extend the technique to polynomial and lower bounds on cost usage in order to combine it with our framework for differential cost analysis.

\myparagraph{Proving symbolic polynomial bounds.} We conclude this section by describing how our algorithm may be adapted to proving symbolic polynomial bounds on the difference in cost usage. Let $p$ be a polynomial function over the set of program variables $V$ and suppose that we want to prove the differential cost assertion
\[ \forall \mathbf{x}\in \initvars.\, \CostSup_{\newsub{\transys}}(\newsub{\loc_0},\mathbf{x}) - \CostInf_{\oldsub{\transys}}(\oldsub{\loc_0},\mathbf{x}) \leq p(\mathbf{x}). \]
In order to use our algorithm to prove such an assertion, one should drop the minimization objective in Step~$4$ and replace every appearance of the constant threshold $t$ in the algorithm description by the polynomial $p$. By doing this and by requiring that the maximal polynomial degree parameter $d$ is greater than or equal to the degree of $p$, we may use our algorithm to prove the desired differential cost assertion.
The only difference in using our algorithm to prove a polynomial bound $p$ as opposed to a concrete threshold value $t$ would occur in Step~$2$, as this change would introduce polynomial $p$ in the differential cost constraint. However, as the polynomial would appear on the right-hand-side of the implication, the conversion in Step~$3$ of the algorithm would still reduce the DiffCost problem to an LP instance and the rest of the algorithm would proceed analogously as in the case of the concrete threshold value $t$.

%% file: exp.tex
\section{Empirical Evaluation}
\label{sec:experiments}

\JEDI{Research Question: Is the potential/anti-potential method effective at getting tight bounds?}

In this section, we evaluate our approach to differential cost analysis by considering the following research question:
\begin{asparadesc}
  \item[Tightness of Differential Thresholds.] Does the simultaneous derivation of potential and anti-potential functions with a threshold value $t$ yield tight differential bounds?
\end{asparadesc}

\JEDI{Bottom-Line Results: 14/19 with tight bounds. Provides some evidence.}

To address this question, we consider $19$ program pairs as benchmarks on which we perform differential cost analysis in Table~\ref{tab:exp}.
In the end, our method computed tight thresholds for 74\% (14/19) of the benchmarks.
There were 2 benchmarks where we failed to compute any threshold, and we discuss the reason behind the failure below.
The run times of our tool (including invariant generation, extraction of constraints, and linear programming) are of the order of a few seconds and suggest that run time is not the bottleneck.

\begin{table}[t]\centering
	\caption{Tightness of differential thresholds.
	For each benchmark, we note the \emph{Tight} differential threshold bound (i.e., the maximal difference in cost usage that can be attained, which is determined manually) and the one we \emph{Computed}. 
	Computed thresholds that are tight and match the maximal difference are given in \textbf{bold}.
	Finally, the \emph{Time} column shows the time (in seconds) taken by our tool to compute the threshold.
	The \xmark{} indicates the cases where we were unable to compute a threshold.
	The programs are drawn the literature on cost analysis and semantic differencing, with the first group of 10 from \citet{GulwaniMC09}, the next 5 from 
	\citet{GulwaniZ10}, and the last 4 from \citet{PartushY14,PartushY13}.
	The program names are the original names used in the works from which the examples are taken, and consist of methods that are between 10 and 20 lines of code.
	The $^\ast$ marks cases where the off-the-shelf invariant generators miss some simple, expected invariants about the loop bounds being satisfied upon entering the loop bodies and so where we slightly strengthened them.}
	\begin{tabular*}{\linewidth}{@{\extracolsep{\fill}}lrr<{\phantom{\textbf{.00}}}r@{}} \toprule
		Benchmark & \multicolumn{2}{c}{Threshold (n)}    & Time (s) \\\cmidrule(lr){2-3}
				  & Tight & \multicolumn{1}{r}{Computed} \\\midrule
		\multicolumn{4}{@{}l}{\underline{Non-Zero Tight Threshold}} \\
		Dis1                              & \textbf{100}  & \textbf{100}  & 2.5 \\
		Dis2                              & \textbf{100}  & \textbf{100}  & 3.4 \\
		NestedMultiple                    & \textbf{100}  & \textbf{100}  & 4.3 \\
		NestedMultipleDep$^\ast$          & \textbf{9900} & \textbf{9900} & 1.5 \\
		NestedSingle                      & \textbf{101}  & \textbf{101}  & 0.9 \\
		SequentialSingle                  & \textbf{100}  & \textbf{100}  & 0.9 \\
		SimpleMultiple                    & \textbf{100}  & \textbf{100}  & 1.5 \\
		SimpleMultipleDep                 & 10000         & 10100         & 1.4 \\
		SimpleSingle                      & \textbf{100}  & \textbf{100}  & 0.7 \\
		SimpleSingle2                     & 100           & 197           & 2.0 \\[0.5ex]
		Ex2 & \textbf{99}  & \multicolumn{1}{r}{\textbf{99.94}} & 2.3 \\
		Ex4 & \textbf{201} & \textbf{201}                       & 0.9 \\
		Ex5 & 100          & \multicolumn{1}{c}{\xmark}         & 2.1 \\
		Ex6 & \textbf{99}  & \multicolumn{1}{r}{\textbf{99.01}} & 1.4 \\
		Ex7 & 1            & \multicolumn{1}{c}{\xmark}         & 1.4 \\[0.5ex]
		\multicolumn{4}{@{}l}{\underline{Zero Tight Threshold}} \\
		ddec                   & 0          & 73896.4         & 0.9 \\
		ddec modified          & \textbf{0} & \textbf{0}   & 0.9 \\
		nested$^\ast$          & \textbf{0} & \textbf{0}   & 2.0 \\
		sum                    & \textbf{0} & \textbf{0.5} & 1.7 \\\bottomrule
	\end{tabular*}
	\label{tab:exp}
\end{table}

\JEDI{Experimental Design: Consider program pairs with complex looping and conditional patterns.}

To evaluate the tightness of the differential thresholds we can derive with simultaneous computation of potentials and anti-potentials, we consider two classes of benchmarks. The first consist of examples that increase the cost (i.e., the tightest possible threshold is non-zero), while the second consist of examples that do not change the cost but have non-trivial cost-preserving changes (e.g., would be non-trivial to align for relational reasoning).

\JEDI{Experimental Setup: Dig into the 3 classes of benchmarks. Complex ones drawn from the literature.}

In the first class of benchmarks, we obtain programs from the single-program cost analysis literature (specifically,~\cite{GulwaniMC09,GulwaniZ10}) to emphasize testing the cost reasoning capability. These programs involve combinations of single or nested loops with conditional branching and non-determinism, and present complex looping patterns. We consider these as representatives of the class of real world examples that involve, for example, iterating through collections with branching or non-deterministic behavior of a function called through an interface. These are code patterns that can often lead to unpredictable resource usage and for which automated differential cost analysis is particularly beneficial. From \citet{GulwaniZ10}, we omit Ex1 and Ex3, as the off-the-shelf tool we use to translate C programs to transition systems (C2fsm~\cite{FeautrierG10}) does not support booleans and pointers.

From each of these programs with interesting resource usage, we produce a revision or program pair for differential cost analysis as follows. For the first program, we make it incur a cost of $1$ for each loop iteration so that the total cost usage corresponds to the loop bound for the program. This choice is to match the original intent of the these benchmarks, as they are drawn from work on loop-bound analysis. In the second program, we pick either a nested loop or an if-branch to incur a cost of $1$. This choice of selecting a nested loop or an if-branch to incur cost makes the differential cost analysis non-trivial (i.e., the second program does not have the same cost as the first nor $0$ cost, and differential cost analysis requires reasoning about different program behavior in each loop iteration).

In the second class of benchmarks, we obtain examples from the semantic differencing literature (specifically, \cite{PartushY14,PartushY13}) to emphasize testing the differential reasoning capability. These are examples of semantically equivalent program pairs, so their relative cost is $0$ on every input. In $2$ out of the $3$ examples, there is a syntactic difference that does not allow program alignment---making it difficult to apply a relational approach. The remaining program pairs from that work either involve boolean variables and pointers, or are straightline programs for which cost analysis is trivial.

For all of the examples, our method computes polynomial cost bounds with the maximal polynomial degree and the maximal number of terms in products in Step~$3$ of our algorithm being $d=K=2$, except for `nested' in which the total cost usage is cubic so we use $d=K=3$. Finally, for each uninitialized program variable we assume that its initial value is in the interval $[1,100]$. In cases when a loop bound is defined by the difference between the variable value and its symbolic initial value, we assume that this difference is in the interval $[1,100]$. We do this to bound initial variable values, so that the difference in cost usage between programs is also bounded and we may use the aforementioned program pairs for a meaningful evaluation of our approach. In Dis2, we in addition assume an initial ordering of variable values to avoid the need for disjunctive reasoning.

\JEDI{Limitations and Threats to Validity: Disjunctive reasoning. Made up revisions.}

\myparagraph{Limitations and Threats to Validity.}
We inspected examples for which our approach is either not able to compute a bound or for which the computed bound is not tight. In particular, we observe that SimpleMultipleDep, SimpleSingle2, Ex5, Ex7 and ddec all require disjunctive reasoning in order to compute tight cost bounds. Disjunctive reasoning is somewhat orthogonal, but it does indicate that disjunctive reasoning is a limitation as-is (as discussed in Section~\ref{sec:algorithm}) and is indeed important for differential cost analysis like other analysis domains.
The program `ddec modified' is a modification of ddec that does not require disjunctive reasoning for cost analysis, and we see that for this example our method computes a tight bound on cost difference.


Another interesting thing to note in Table~\ref{tab:exp} is the small imprecision in bounds computed for Ex2, Ex4 and `sum'. Observe that the computed bounds for these examples slightly exceed the maximal difference in cost usage, with the difference being smaller than $1$ (as a result of real-valued linear programming). Nevertheless, since we consider programs with integer variables and costs, the computed bound is tight.

A natural threat to validity is whether these benchmarks are representative. We have attempted to mitigate this threat by considering two classes of benchmarks focusing on different aspects and considering looping patterns for common code like iterating over collections. There are few works that study differential cost analysis, and~\cite{QuG019,CicekQBG019} experimentally evaluate their approaches on functional programs that manipulate list-like objects. Since we consider differential cost analysis in general imperative programs for which, to the best of our knowledge, there is no existing benchmark set, we created our own benchmark set by collecting example imperative programs from the literature.

\myparagraph{Implementation.}
In order to empirically evaluate our approach, we implemented a prototype tool which takes C programs as input and uses C2fsm~\cite{FeautrierG10} to translate them to equivalent transition systems. We note that C2fsm supports a slightly out-of-date dialect of C that does not allow boolean data types or constructs such as structs or pointers. However, it supports general numerical data types and control-flow constructs, making it is sufficient for our evaluation. For invariant generation we use Aspic~\cite{FeautrierG10} and Sting~\cite{SankaranarayananSM04}, and we use Gurobi~\cite{gurobi} to solve linear programs. All experiments were run on Ubuntu with an Intel(R) Core(TM) i5-8250U CPU at 1.60GHz and with 16GB of RAM.

%% file: tightness.tex
\section{Precision Guarantees on Bounds for a Single Program}\label{sec:tight}

As noted in Section~\ref{sec:overview}, while the motivation for simultaneously computing potential and anti-potential functions is to address the differential cost analysis problem, an additional consequence of our approach is that it suggests a way to compute cost bounds with \emph{precison guarantees}.  We now show how our approach can be adapted to compute upper and lower bounds on cost incurred in a single program with guarantees on the precision of computed bounds.
And we have not seen other cost analyses that provide such guarantees on the quality of computed cost bounds.
    
Let $\transys = (\locs,\vars,\trans,\locinit,\initvars)$ be a terminating transition system that models a program whose cost usage we wish to analyze. We may compute bounds on the program's cost usage by naturally adapting our algorithm presented in Section~\ref{sec:algorithm} to compute (1)~a PF $\potential$ in $\transys$, (2)~an anti-PF $\antipotential$ in $\transys$, and (3)~a value $p$ that satisfies
\[ \forall\mathbf{x}\in\Theta_0.\, \potential(\locinit,\mathbf{x}) - \antipotential(\locinit,\mathbf{x}) \leq p. \]
The resulting algorithm simultaneously computes $\potential$, $\antipotential$ and $p$ by reduction to an LP instance that minimizes $p$. The computed value of $p$ is a {\em bound on precision} of both the computed upper bound $\potential$ and the lower bound $\antipotential$ on cost incurred in $\transys$. The following theorem proves that our algorithm is sound and that it indeed provides precision guarantees. The proof can be found in Appendix~\ref{app:proofs}.

\begin{theorem}\label{thm:precision}
	Let $\transys$ be a terminating transition systems. Suppose that $\potential$ is a PF and that $\antipotential$ is an anti-PF in $\transys$. Then, for each initial variable valuation $\mathbf{x}\in\initvars$, we have that
	\begin{equation*}
	\CostSup_{\transys}(\locinit,\mathbf{x}) - \CostInf_{\transys}(\locinit,\mathbf{x}) \leq \potential(\locinit,\mathbf{x}) - \antipotential(\locinit,\mathbf{x}).
	\end{equation*}
	In particular, if $p$ satisfies $\potential(\locinit,\mathbf{x}) - \antipotential(\locinit,\mathbf{x}) \leq p$ for each $\mathbf{x}\in\initvars$, then for any run $\rho$ that starts in some initial state $(\locinit,\mathbf{x})$ with $\mathbf{x}\in\Theta_0$ we have $ 0\leq \Cost_{\transys}(\rho)-\antipotential(\locinit,\mathbf{x}) \leq p$ and $ 0\leq \potential(\locinit,\mathbf{x}) - \Cost_{\transys}(\rho) \leq p$. Hence, $p$ is a bound on the precision of the upper cost bound defined by $\potential$ and the lower cost bound defined by $\antipotential$ in $\transys$.
\end{theorem} 

We note that the lack of disjunctive reasoning in our method could result in weaker bounds for programs in which disjunctive reasoning for computing tight cost bounds is needed, when compared to some approaches
that do compute them~\cite{GulwaniZ10,Carbonneaux0S15,SinnZV17}.
However, as new techniques are developed that compute tighter bounds, our approach may offer a way to get both tight bounds and precision guarantees.

%% file: relatedwork.tex
\section{Related Work}\label{sec:relatedwork}

\noindent{\em Differential cost analysis.} The existing works on differential cost analysis consider functional programs and propose relational type and effect systems to reason about relative cost between programs~\cite{CicekBG0H17,RadicekBG0Z18,CicekQBG019}, and the relational type and effect system of~\cite{QuG019} allows reasoning about functional programs with mutable arrays. The relational analysis is done by syntactically aligning two programs and considering relational types that capture the difference in cost incurred in two programs. Once the alignment is no longer possible, these works use unary types to allow cost analysis in a single program.
%
As discussed in Section~\ref{sec:intro}, our approach offers an alternative with different trade-offs.
For example, when given a pair of syntactically similar programs, relational type systems naturally exploit this similarity through syntactic alignment of programs, which may simplify differential cost analysis~\cite{CicekBG0H17}. And in contrast to the approaches based on relational type systems, our method is not compositional. It reduces differential cost analysis to solving a linear program, which is a global optimization problem whose constraints may depend on all parts of the program. An interesting direction of future work is to extend our method to generate method summaries~\cite{GulwaniSV08} that can express quantitative specifications on cost (to make it more compositional).

The Infer Static Analyzer~\cite{DistefanoFLO19} performs a kind of differential cost analysis on industrial size codebases using a worst-case execution time (WCET) analysis~\cite{Bygde1789}. It considers C or Java programs and performs unary cost analysis on each program to compute two polynomial cost upper bounds, which are then compared. A warning is raised if there is a jump in polynomial degree of incurred cost~\cite{CicekBCD20}. Thus, their method detects increases in resource usage only when there is an increase in polynomial degree.
However, it is impressively scalable and in many cases successfully detects increases in cost polynomial degrees. Given that Infer also focuses on imperative programming languages, it may be possible to couple it with our approach in order to obtain a tool which scales to large codebases, but which allows a sound and precise differential cost analysis for parts of programs that are deemed performance critical. One possible strategy is
to use our approach for precise amortized reasoning locally within a global worst-case reasoning framework~\cite{DBLP:conf/sas/LuCT21}.

\myparagraph{Static (unary) cost analysis.} Static cost analysis for single programs is a classical and well studied problem. There are many existing methods, and most works focus on computing upper bounds on cost usage with techniques based on amortized analysis~\cite{HoffmannAH11,HoffmannH10,HoffmannDW17,Carbonneaux0S15,0002AH12}, type systems~\cite{HofmannJ03,AvanziniL17,LagoG11,LagoP13}, term-rewriting and abstract interpretation~\cite{BrockschmidtE0F16,GulwaniZ10,Gulwani09,GulwaniMC09}, ranking functions~\cite{AliasDFG10}, invariant generation~\cite{KincaidBBR17} or the analysis of abstract program models~\cite{SinnZV14,ZulegerGSV11,SinnZV17}. In particular, \cite{HoffmannAH11,HoffmannH10,HoffmannDW17,Carbonneaux0S15,0002AH12} also compute potential functions for amortized analysis by reduction to linear programming. The approach of \citet{KincaidBBR17} that is based on invariant generation can be used to obtain both upper and lower bounds on cost usage with bounds that may involve piecewise-linear operators max and min. Thus, similarly to \citet{Carbonneaux0S15}, it would be interesting to consider the possibility of combining it with our framework towards obtaining piecewise-polynomial bounds for differential cost analysis. Computing lower bounds on program run time has been considered~\cite{FrohnNBG20}. \citet{NgoDFH17} proposes a type system for verifying that a functional program has constant resource usage, which has important implications in security and prevents leakage of confidential information under side-channel attacks that exploit non-constant run times or energy consumption. Their type system defines a potential-like function whose values before and after evaluating an expression differ exactly by the cost of the evaluation. This corresponds to strengthening the \emph{sufficiency} preservation condition of PFs and the \emph{insufficiency} preservation condition of anti-PFs by imposing the strict equality ``='' sign. By relaxing type judgements with ``$
\geq$'' or ``$\leq$'' signs, the type system is relaxed to a type system for computing upper or lower bounds on cost usage and type checking would correspond to computing a PF or an anti-PF, respectively. Computing upper and lower bounds on the expected cost usage in probabilistic programs has also been considered~\cite{NgoC018,Wang0GCQS19,AvanziniMS20}. In particular, our technique for translating the defining conditions of PFs and anti-PFs into linear constraints has similarities to that of \citet{Wang0GCQS19}, which defines a probabilistic variant of potential functions called upper and lower cost supermartingales and uses Handelman's theorem for their computation. Handelman's theorem and other results from algebraic geometry are used by~\citet{ChatterjeeFG16} for computing probability~$1$ termination certificates and bounds on termination time.
While~\citet{NgoC018} and~\citet{Wang0GCQS19} consider computation of both upper and lower cost bounds, they study unary cost analysis. If one attempted to directly adapt these methods to differential cost analysis by computing two cost bounds and comparing them, this would be precisely the na\"ive approach discussed in Section~\ref{sec:intro} that is problematic for precision. This limitation of the na\"ive approach was also pointed out in prior work on differential cost analysis and was used to motivate the use of relational type systems~\cite{CicekBG0H17}. In contrast, we compute upper and lower cost bounds for a pair of programs \emph{together with a threshold value} to address the differential cost analysis problem; it is this simultaneous computation that side-steps the limitations of the na\"ive approach.



\myparagraph{Constraint solving-based program analysis.} Constraint solving-based techniques are a classical approach to program analysis~\cite{GulwaniSV08}, that have been used for multiple static analyses including the synthesis of ranking functions for termination analysis~\cite{ColonS01,BradleyMS05,AliasDFG10}, proving non-termination~\cite{LarrazNORR14,ChatterjeeG0Z21}, invariant generation~\cite{ColonSS03,Chatterjee0GG20}, reachability~\cite{AsadiC0GM21}, as well as several methods for cost analysis that we discussed above.

\section{Conclusion}\label{sec:conclusion}

We present a novel approach to differential cost analysis for imperative programs that uses potential and anti-potential functions to reason about the difference in incurred cost.
A threshold value on the maximal difference in cost between two program versions is computed simultaneously with a potential function that provides an upper bound on the cost incurred in the new version and an anti-potential function that provides a lower bound on the cost incurred in the old one.
This dual potential-based method side steps the need for and limitations of version alignment and offers a complimentary approach to ones based on relational reasoning.
To automatically derive the potential function, the anti-potential function, and the threshold for differential cost analysis, we employ a constraint solving-based approach that has the benefit of using off-the-shelf invariant generators and linear program solvers, as well as supporting optimization of the threshold bound.



%% file: app.tex
\begin{center}
	{\Large Appendix}
\end{center}

\section{Transition System for the Running Example}\label{app:transitionsystem}

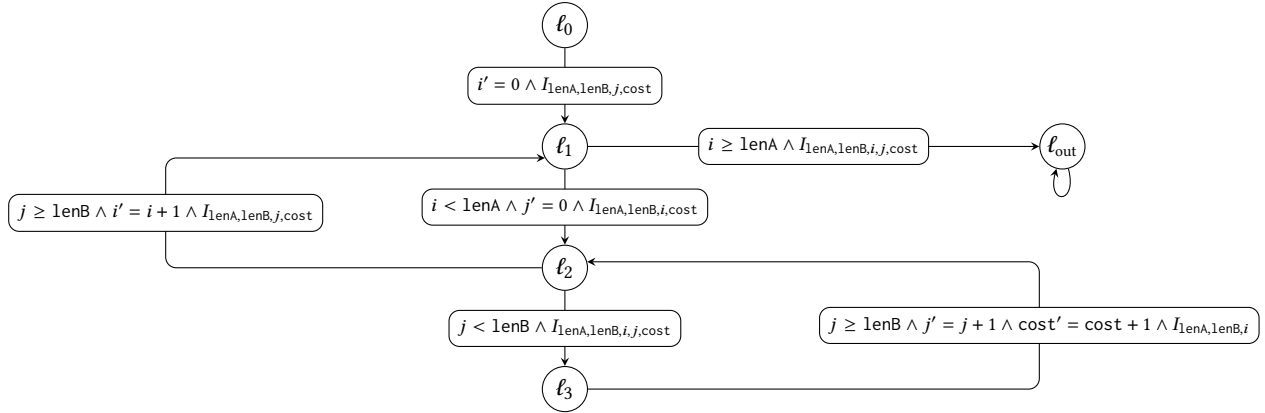
\begin{figure*}[t]
	\centering
	\begin{tikzpicture}
	\node[ran] (start) at (0,0)  {$\loc_0$};
	\node[ran, below = 1 cm of start] (loop1) {$ \loc_1 $};
	\node[ran, right = 6cm of loop1] (term) {$ \locterm $};
	\node[ran, below = 1 cm of loop1] (loop2) {$ \loc_2 $};
	\node[ran, below = 1 cm of loop2] (calltof) {$ \loc_3 $};
	
	\draw[tran] (start) to node[font=\scriptsize,draw, fill=white, 
	rectangle,pos=0.5] {$i' = 0  \land I_{\lenA,\lenB,j,\cost}$} (loop1);
	\draw[tran] (loop1) to node[font=\scriptsize,draw, fill=white, 
	rectangle,pos=0.5] {$i \geq \lenA \land I_{\lenA,\lenB,i,j,\cost}$} (term);
	\draw[tran] (loop1) to node[font=\scriptsize,draw, fill=white, 
	rectangle,pos=0.5] {$i < \lenA \land j' = 0  \land I_{\lenA,\lenB,i,\cost}$} (loop2);
	\node[left = 5cm of loop2, circle, minimum size = 3mm] (dum) {};
	\draw[tran, rounded corners] (loop2) -- (dum.east) -- node[font=\scriptsize,draw, fill=white, 
	rectangle,pos=0.5] {$j\geq \lenB \land i'=i+1 \land I_{\lenA,\lenB,j,\cost}$} (dum.east|-loop1.210) -- (loop1.210);
	\draw[tran] (loop2) to node[font=\scriptsize,draw, fill=white, 
	rectangle,pos=0.5] {$j < \lenB \land I_{\lenA,\lenB,i,j,\cost}$} (calltof);
	\node[right = 6cm of calltof, circle, minimum size = 3mm] (dum2) {};
	\draw[tran, rounded corners] (calltof) -- (dum2.west) -- node[font=\scriptsize,draw, fill=white, 
	rectangle,pos=0.5] {$j\geq \lenB \land j'=j+1 \land \cost' = \cost + 1 \land I_{\lenA,\lenB,i}$} (dum2.west|-loop2.15) -- (loop2.15);
	\draw[tran, loop below] (term) to  (term);
	\end{tikzpicture}
	\caption{The transition system that models the old version of the procedure \lstinline{join} in Fig.~\ref{fig:running} left. Transition updates are presented as equalities that assign new values to each program variable, with new variable values denoted by the primed notation that is standard in program analysis. For a vector of variables $\tilde{V}$, we use $I_{\tilde{V}}$ to denote the logical formula $\bigwedge_{v\in\tilde{V}}v'=v$. This notation is used for readability. A transition system for the new version of the procedure \lstinline{join} in Fig.~\ref{fig:running} right is almost identical, with the exception that each appearance of $\lenA$ is replaced by $\lenB$ and vice-versa, and the update of the transition from $\loc_3$ to $\loc_2$ contains the term $\cost'=\cost + 2$ instead of $\cost'=\cost + 1$.}
	\label{fig:runningtransys}
\end{figure*}

The transition system that models the old version of the procedure \lstinline{join} in Fig.~\ref{fig:running} left is presented in Fig.~\ref{fig:runningtransys}. It contains $5$ program locations $L=\{\loc_0,\loc_1,\loc_2,\loc_3,\locterm\}$, where $\loc_0$ represents the start of the method, $\loc_1$ and $\loc_2$ are the heads of the outer and the inner loop, $\loc_3$ is the location of the call to the operator \lstinline{f} and $\locterm$ is the terminal location. The variable set is given by $\vars = \{\lenA,\lenB,i,j,\cost\}$. Note that elements of the arrays $\arrA$ and $\arrB$ are not included into the variable set. This is because they do not contribute to the cost usage, and their removal can be automated through program slicing. Locations are depicted by labeled circles, and transitions are depicted by arrows between the locations. Transition guards and updates are presented in boxes along the transitions.

A transition system for the new version of the procedure \lstinline{join} in Fig.~\ref{fig:running} right is almost identical, with the exception that each appearance of $\lenA$ is replaced by $\lenB$ and vice-versa, and the update of the transition from $\loc_3$ to $\loc_2$ contains the term $\cost'=\cost + 2$ instead of $\cost'=\cost + 1$.

\section{Theorem Proofs}\label{app:proofs}

\begin{manualtheorem}{4.1}
	Let $\transys$ be a transition system that is terminating. If $\potential$ is a PF in $\transys$, then for any reachable state $(\loc,\mathbf{x})$ in $\transys$ we have
	\[ \potential(\loc,\mathbf{x}) \geq \CostSup_{\transys}(\loc,\mathbf{x}). \]
	If $\antipotential$ is an anti-PF in $\transys$, then for any reachable state $(\loc,\mathbf{x})$ in $\transys$ we have
	\[ \antipotential(\loc,\mathbf{x}) \leq \CostInf_{\transys}(\loc,\mathbf{x}). \]
\end{manualtheorem}

\begin{proof}
	Suppose first that $\potential$ is a PF in $\transys$ and that $(\loc,\mathbf{x})$ is a reachable state in $\transys$. We need to show that $\potential(\loc,\mathbf{x}) \geq \CostSup_{\transys}(\loc,\mathbf{x})$. Since $\CostSup_{\transys}(\loc,\mathbf{x})=\sup \{\Cost_{\transys}(\rho)\mid \rho\in \Run(\loc,\mathbf{x})\}$, it suffices to prove that $\potential(\loc,\mathbf{x}) \geq \Cost_{\transys}(\rho)$ for each $\rho\in \Run(\loc,\mathbf{x})$. On the other hand, since $\transys$ is terminating, we know that each run in $\Run(\loc,\mathbf{x})$ terminates in finitely many steps. Hence, we may prove that $\potential(\loc,\mathbf{x}) \geq \Cost_{\transys}(\rho)$ for each $\rho\in \Run(\loc,\mathbf{x})$ by induction on the number of steps in which $\rho$ terminates, which we denote by $\text{len}(\rho)$.
	\begin{description}
		\item[Base case: $\text{len}(\rho)=0$.] If $\text{len}(\rho)=0$, then $(\loc,\mathbf{x})$ is a terminal state and $\Cost_{\transys}(\rho)=\mathbf{x}[\cost]-\mathbf{x}[\cost]=0$. On the other hand, by the Potential on termination we have $\potential(\loc,\mathbf{x})\geq 0$. Hence, $\potential(\loc,\mathbf{x}) \geq \Cost_{\transys}(\rho)$ and the claim holds for the base case.
		
		\item[Induction hypothesis.] Suppose that $k\in\mathbb{N}_{\geq 0}$ and that, for each $\rho\in \Run(\loc,\mathbf{x})$ with $\text{len}(\rho)\leq k$, we have $\potential(\loc,\mathbf{x}) \geq \Cost_{\transys}(\rho)$.
		
		\item[Induction step: proof for $k+1$.] Let $\rho\in \Run(\loc,\mathbf{x})$ with $\text{len}(\rho) = k+1$. We prove that $\potential(\loc,\mathbf{x}) \geq \Cost_{\transys}(\rho)$. Decompose the run $\rho$ into $(\loc,\mathbf{x}),(\loc_1,\mathbf{x}_1),\rho_1$, where $(\loc_1,\mathbf{x}_1)$ is the second state along $\rho$ and $\rho_1$ is a suffix of $\rho$ of length $k$. Then, we have
		\begin{equation}
		\begin{split}
		\Cost_{\transys}(\rho) &= \mathbf{x}[\cost] - \mathbf{x}_1[\cost] + \Cost_{\transys}(\rho_1) \\
		&\leq \potential(\loc,\mathbf{x}) - \potential(\loc_1,\mathbf{x}_1) + \Cost_{\transys}(\rho_1) \\
		&\leq \potential(\loc,\mathbf{x}) - \potential(\loc_1,\mathbf{x}_1) + \potential(\loc_1,\mathbf{x}_1) \\
		&= \potential(\loc,\mathbf{x}),
		\end{split}
		\end{equation}
		where the first inequality holds by the Sufficient resource preservation condition, and the second inequality holds by induction hypothesis. This concludes the proof by induction.
	\end{description}

	We now prove the second part of the theorem claim. Suppose that $\antipotential$ is an anti-PF in $\transys$ and that $(\loc,\mathbf{x})$ is a reachable state in $\transys$. We need to show that $\antipotential(\loc,\mathbf{x}) \leq \CostInf_{\transys}(\loc,\mathbf{x})$. Since $\CostInf_{\transys}(\loc,\mathbf{x})=\inf \{\Cost_{\transys}(\rho)\mid \rho\in \Run(\loc,\mathbf{x})\}$, it suffices to prove that $\antipotential(\loc,\mathbf{x}) \leq \Cost_{\transys}(\rho)$ for each $\rho\in \Run(\loc,\mathbf{x})$. Hence, again, since $\transys$ is terminating we may prove that $\antipotential(\loc,\mathbf{x}) \leq \Cost_{\transys}(\rho)$ for each $\rho\in \Run(\loc,\mathbf{x})$ by induction on the length of run $\rho$, i.e.~$\text{len}(\rho)$.
	\begin{description}
		\item[Base case: $\text{len}(\rho)=0$.] If $\text{len}(\rho)=0$, then $(\loc,\mathbf{x})$ is a terminal state and $\Cost_{\transys}(\rho)=\mathbf{x}[\cost]-\mathbf{x}[\cost]=0$. On the other hand, by the Anti-potential on termination we have $\antipotential(\loc,\mathbf{x})\leq 0$. Hence, $\antipotential(\loc,\mathbf{x}) \leq \Cost_{\transys}(\rho)$ and the claim holds for the base case.
		
		\item[Induction hypothesis.] Suppose that $k\in\mathbb{N}_{\geq 0}$ and that, for each $\rho\in \Run(\loc,\mathbf{x})$ with $\text{len}(\rho)\leq k$, we have $\antipotential(\loc,\mathbf{x}) \leq \Cost_{\transys}(\rho)$.
		
		\item[Induction step: proof for $k+1$.] Let $\rho\in \Run(\loc,\mathbf{x})$ with $\text{len}(\rho) = k+1$. We prove that $\antipotential(\loc,\mathbf{x}) \leq \Cost_{\transys}(\rho)$. Decompose the run $\rho$ into $(\loc,\mathbf{x}),(\loc_1,\mathbf{x}_1),\rho_1$, where $(\loc_1,\mathbf{x}_1)$ is the second state along $\rho$ and $\rho_1$ is a suffix of $\rho$ of length $k$. Then, we have
		\begin{equation}
		\begin{split}
		\Cost_{\transys}(\rho) &= \mathbf{x}[\cost] - \mathbf{x}_1[\cost] + \Cost_{\transys}(\rho_1) \\
		&\geq \antipotential(\loc,\mathbf{x}) - \antipotential(\loc_1,\mathbf{x}_1) + \Cost_{\transys}(\rho_1) \\
		&\geq \antipotential(\loc,\mathbf{x}) - \antipotential(\loc_1,\mathbf{x}_1) + \antipotential(\loc_1,\mathbf{x}_1) \\
		&= \antipotential(\loc,\mathbf{x}),
		\end{split}
		\end{equation}
		where the first inequality holds by the Insufficient-resource preservation condition, and the second inequality holds by induction hypothesis. This concludes the proof by induction.
	\end{description}
\end{proof}

\begin{manualtheorem}{4.2}[PFs and anti-PFs for DiffCost]
	Let $\newsub{\transys}$ and $\oldsub{\transys}$ be two terminating transition systems. Suppose that $\newsub{\potential}$ is a PF in $\newsub{\transys}$ and that $\oldsub{\antipotential}$ is an anti-PF in $\oldsub{\transys}$. Then, for each initial variable valuation $\mathbf{x}\in\initvars$, we have that
	\begin{equation*}
	\begin{split}
	&\CostSup_{\newsub{\transys}}(\newsub{\loc_0},\mathbf{x}) - \CostInf_{\oldsub{\transys}}(\oldsub{\loc_0},\mathbf{x}) \\
	&\leq \newsub{\potential}(\newsub{\locinit},\mathbf{x}) - \oldsub{\antipotential}(\oldsub{\locinit},\mathbf{x}).
	\end{split}
	\end{equation*}
	In particular, if $t$ satisfies $\newsub{\potential}(\newsub{\locinit},\mathbf{x}) - 
	\oldsub{\antipotential}(\oldsub{\locinit},\mathbf{x}) \leq t$ for each $\mathbf{x}\in\initvars$, then $t$ is a threshold for the DiffCost problem.
	
	Conversely, if $t\in\mathbb{Z}$ is a threshold for the DiffCost problem, then there exist a PF $\newsub{\potential}$ in $\newsub{\transys}$ and an anti-PF $\oldsub{\antipotential}$ in $\oldsub{\transys}$ such that $\newsub{\potential}(\newsub{\locinit},\mathbf{x}) - 
	\oldsub{\antipotential}(\oldsub{\locinit},\mathbf{x}) \leq t$ holds for each $\mathbf{x}\in\initvars$.
\end{manualtheorem}

\begin{proof}
	Let $\newsub{\potential}$ be a PF in $\newsub{\transys}$, $\oldsub{\antipotential}$ be an anti-PF in $\oldsub{\transys}$ and $\mathbf{x}\in\initvars$. Then, by Theorem~\ref{thm:potential}, it follows that
	\[ \CostSup_{\newsub{\transys}}(\newsub{\loc_0},\mathbf{x}) \leq \newsub{\potential}(\newsub{\locinit},\mathbf{x}) \]
	and
	\[ \CostInf_{\oldsub{\transys}}(\oldsub{\loc_0},\mathbf{x}) \geq \oldsub{\antipotential}(\oldsub{\locinit},\mathbf{x}). \]
	Combining the two inequalities, we conclude that
	\begin{equation*}
	\begin{split}
	&\CostSup_{\newsub{\transys}}(\newsub{\loc_0},\mathbf{x}) - \CostInf_{\oldsub{\transys}}(\oldsub{\loc_0},\mathbf{x}) \\
	&\leq \newsub{\potential}(\newsub{\locinit},\mathbf{x}) - \oldsub{\antipotential}(\oldsub{\locinit},\mathbf{x}),
	\end{split}
	\end{equation*}
	as desired.
	
	To prove the second claim of the theorem, suppose that $t\in\mathbb{Z}$ is a threshold for the DiffCost problem. Define a map $\newsub{\potential}$ that to each reachable state $(\loc,\mathbf{x})$ in $\newsub{\transys}$ assigns
	\[ \newsub{\potential}(\loc,\mathbf{x}) := \CostSup_{\newsub{\transys}}(\loc,\mathbf{x}). \]
	Next, define a map $\oldsub{\antipotential}$ that to each reachable state $(\loc,\mathbf{x})$ in $\oldsub{\transys}$ assigns
	\[ \oldsub{\antipotential}(\loc,\mathbf{x}) := \CostInf_{\oldsub{\transys}}(\loc,\mathbf{x}). \]
	By definition of the threshold value for the DiffCost problem, we know that $\newsub{\potential}(\newsub{\locinit},\mathbf{x}) - 
	\oldsub{\antipotential}(\oldsub{\locinit},\mathbf{x}) \leq t$ holds for each $\mathbf{x}\in\initvars$. Hence, we are left to prove that $\newsub{\potential}$ is a PF in $\newsub{\transys}$ and that $\oldsub{\antipotential}$ is an anti-PF in $\oldsub{\transys}$.
	
	First, we prove that $\newsub{\potential}$ is a PF in $\newsub{\transys}$ by showing that it satisfies the Sufficient resource preservation and the Potential on termination conditions:
	\begin{description}
		\item[Sufficient resource preservation.] Let $(\loc,\mathbf{x})$ be a reachable state in $\newsub{\transys}$ and $(\loc',\mathbf{x}')$ be a successor state of $(\loc,\mathbf{x})$. We need to show that 
		\begin{equation*}
		\newsub{\potential}(\loc,\mathbf{x}) - \newsub{\potential}(\loc',\mathbf{x}') \geq \mathbf{x}'[\cost] - \mathbf{x}[\cost].
		\end{equation*}
		Since we have $\newsub{\potential}(\loc,\mathbf{x}) - \newsub{\potential}(\loc',\mathbf{x}') = \newsub{\potential}(\loc,\mathbf{x}) - \CostSup_{\newsub{\transys}}(\loc',\mathbf{x}')$, it suffices to show that $\newsub{\potential}(\loc,\mathbf{x}) -\Cost_{\newsub{\transys}}(\rho') \geq \mathbf{x}'[\cost] - \mathbf{x}[\cost]$ for each run $\rho'\in\Run(\loc',\mathbf{x}')$.
		
		To prove this, fix a run $\rho'\in\Run(\loc',\mathbf{x}')$ and define $\rho\in\Run(\loc,\mathbf{x})$ via $\rho = (\loc,\mathbf{x}),\rho'$. This is a run in $\newsub{\transys}$ as $(\loc'\mathbf{x}')$ is a successor of $(\loc,\mathbf{x})$. Moreover, we have that $\Cost_{\newsub{\transys}}(\rho)-\Cost_{\newsub{\transys}}(\rho')=\mathbf{x}'[\cost] - \mathbf{x}[\cost]$. Hence,
		\begin{equation*}
		\begin{split}
		\mathbf{x}'[\cost] - &\mathbf{x}[\cost] = \Cost_{\newsub{\transys}}(\rho)-\Cost_{\newsub{\transys}}(\rho') \\
		&\leq \CostSup_{\newsub{\transys}}(\loc,\mathbf{x}) - \Cost_{\newsub{\transys}}(\rho') \\
		&= \newsub{\potential}(\loc,\mathbf{x}) -\Cost_{\newsub{\transys}}(\rho').
		\end{split}
		\end{equation*}
		Since the run $\rho'\in\Run(\loc',\mathbf{x}')$ was arbitrary, we conclude that $\newsub{\potential}(\loc,\mathbf{x}) - \newsub{\potential}(\loc',\mathbf{x}') \geq \mathbf{x}'[\cost] - \mathbf{x}[\cost]$.
		\item[Potential on termination.] If $(\loc,\mathbf{x})$ is a reachable terminal state, then $\newsub{\potential}(\loc,\mathbf{x})=\CostSup_{\newsub{\transys}}(\loc,\mathbf{x})= 0$ since for any run $\rho\in \Run(\loc,\mathbf{x})$ we have $\Cost_{\newsub{\transys}}(\rho)=0$.
	\end{description}
	
	Second, we prove that $\oldsub{\antipotential}$ is an anti-PF in $\oldsub{\transys}$ by showing that it satisfies the Insufficient resource preservation and the Anti-potential on termination conditions:
	\begin{description}
		\item[Insufficient resource preservation.] Let $(\loc,\mathbf{x})$ be a reachable state in $\oldsub{\transys}$ and $(\loc',\mathbf{x}')$ be a successor state of $(\loc,\mathbf{x})$. We need to show that 
		\begin{equation*}
		\oldsub{\antipotential}(\loc,\mathbf{x}) - \oldsub{\antipotential}(\loc',\mathbf{x}') \leq \mathbf{x}'[\cost] - \mathbf{x}[\cost].
		\end{equation*}
		Since we have $\oldsub{\antipotential}(\loc,\mathbf{x}) - \oldsub{\antipotential}(\loc',\mathbf{x}') = \oldsub{\antipotential}(\loc,\mathbf{x}) - \CostInf_{\oldsub{\transys}}(\loc',\mathbf{x}')$, it suffices to show that $\oldsub{\antipotential}(\loc,\mathbf{x}) -\Cost_{\oldsub{\transys}}(\rho') \leq \mathbf{x}'[\cost] - \mathbf{x}[\cost]$ for each run $\rho'\in\Run(\loc',\mathbf{x}')$.
		
		To prove this, fix a run $\rho'\in\Run(\loc',\mathbf{x}')$ and define $\rho\in\Run(\loc,\mathbf{x})$ via $\rho = (\loc,\mathbf{x}),\rho'$. This is a run in $\oldsub{\transys}$ as $(\loc'\mathbf{x}')$ is a successor of $(\loc,\mathbf{x})$. Moreover, we have that $\Cost_{\oldsub{\transys}}(\rho)-\Cost_{\oldsub{\transys}}(\rho')=\mathbf{x}'[\cost] - \mathbf{x}[\cost]$. Hence,
		\begin{equation*}
		\begin{split}
		\mathbf{x}'[\cost] - &\mathbf{x}[\cost] = \Cost_{\oldsub{\transys}}(\rho)-\Cost_{\oldsub{\transys}}(\rho') \\
		&\geq \CostInf_{\oldsub{\transys}}(\loc,\mathbf{x}) - \Cost_{\oldsub{\transys}}(\rho') \\
		&= \oldsub{\antipotential}(\loc,\mathbf{x}) -\Cost_{\oldsub{\transys}}(\rho').
		\end{split}
		\end{equation*}
		Since the run $\rho'\in\Run(\loc',\mathbf{x}')$ was arbitrary, we conclude that $\oldsub{\antipotential}(\loc,\mathbf{x}) - \oldsub{\antipotential}(\loc',\mathbf{x}') \leq \mathbf{x}'[\cost] - \mathbf{x}[\cost]$.
		\item[Potential on termination.] If $(\loc,\mathbf{x})$ is a reachable terminal state, then $\oldsub{\antipotential}(\loc,\mathbf{x})=\CostInf_{\oldsub{\transys}}(\loc,\mathbf{x})= 0$ since for any run $\rho\in \Run(\loc,\mathbf{x})$ we have $\Cost_{\oldsub{\transys}}(\rho)=0$.
	\end{description}
\end{proof}

\begin{manualtheorem}{4.3}[Refuting a threshold]
	Let $\newsub{\transys}$ and $\oldsub{\transys}$ be two terminating transition systems. Suppose that $\newsub{\antipotential}$ is an anti-PF in $\newsub{\transys}$, and that $\oldsub{\potential}$ is a PF in $\oldsub{\transys}$. Then, for each initial variable valuation $\mathbf{x}\in\initvars$, we have that
	\begin{equation*}
	\begin{split}
	&\CostInf_{\newsub{\transys}}(\newsub{\loc_0},\mathbf{x}) - \CostSup_{\oldsub{\transys}}(\oldsub{\loc_0},\mathbf{x}) \\
	&\geq \newsub{\antipotential}(\newsub{\locinit},\mathbf{x}) - \oldsub{\potential}(\oldsub{\locinit},\mathbf{x}).
	\end{split}
	\end{equation*}
	In particular, if $t\in\mathbb{Z}$ satisfies $\newsub{\antipotential}(\newsub{\locinit},\mathbf{x}) - \oldsub{\potential}(\oldsub{\locinit},\mathbf{x}) > t$ for some $\mathbf{x}\in\initvars$, then $t$ is {\em not} a threshold for the DiffCost problem.
	
	Conversely, if $t$ is not a threshold for the DiffCost problem and if $\newsub{\transys}$ and $\oldsub{\transys}$ are induced by deterministic programs, then there exist an anti-PF $\newsub{\antipotential}$ in $\newsub{\transys}$ and a PF $\oldsub{\potential}$ in $\oldsub{\transys}$ such that $\newsub{\antipotential}(\newsub{\locinit},\mathbf{x}) - \oldsub{\potential}(\oldsub{\locinit},\mathbf{x}) > t$ for at least one $\mathbf{x}\in\initvars$.
\end{manualtheorem}

\begin{proof}
	Let $\newsub{\antipotential}$ be a anti-PF in $\newsub{\transys}$, $\oldsub{\potential}$ be a PF in $\oldsub{\transys}$ and $\mathbf{x}\in\initvars$. Then, by Theorem~\ref{thm:potential}, it follows that
	\[ \CostInf_{\newsub{\transys}}(\newsub{\loc_0},\mathbf{x}) \geq \newsub{\antipotential}(\newsub{\locinit},\mathbf{x}) \]
	and
	\[ \CostSup_{\oldsub{\transys}}(\oldsub{\loc_0},\mathbf{x}) \leq \oldsub{\potential}(\oldsub{\locinit},\mathbf{x}). \]
	Combining the two inequalities, we conclude that
	\begin{equation*}
	\begin{split}
	&\CostInf_{\newsub{\transys}}(\newsub{\loc_0},\mathbf{x}) - \CostSup_{\oldsub{\transys}}(\oldsub{\loc_0},\mathbf{x}) \\
	&\geq \newsub{\antipotential}(\newsub{\locinit},\mathbf{x}) - \oldsub{\potential}(\oldsub{\locinit},\mathbf{x}),
	\end{split}
	\end{equation*}
	as desired. If $t\in\mathbb{Z}$ satisfies $\newsub{\antipotential}(\newsub{\locinit},\mathbf{x}) - \oldsub{\potential}(\oldsub{\locinit},\mathbf{x}) > t$ for some $\mathbf{x}\in\initvars$, then
	\begin{equation*}
	\begin{split}
	&\CostSup_{\newsub{\transys}}(\newsub{\loc_0},\mathbf{x}) - \CostInf_{\oldsub{\transys}}(\oldsub{\loc_0},\mathbf{x}) \\
	&\geq \CostInf_{\newsub{\transys}}(\newsub{\loc_0},\mathbf{x}) - \CostSup_{\oldsub{\transys}}(\oldsub{\loc_0},\mathbf{x}) \\
	&\geq \newsub{\antipotential}(\newsub{\locinit},\mathbf{x}) - \oldsub{\potential}(\oldsub{\locinit},\mathbf{x}) > t,
	\end{split}
	\end{equation*}
	and so $t$ is {\em not} a threshold for the DiffCost problem.
	
	To prove the second claim of the theorem, suppose that $t$ is not a threshold for the DiffCost problem and that $\newsub{\transys}$ and $\oldsub{\transys}$ are induced by deterministic programs. Define a map $\newsub{\antipotential}$ that to each reachable state $(\loc,\mathbf{x})$ in $\newsub{\transys}$ assigns
	\[ \newsub{\potential}(\loc,\mathbf{x}) := \CostInf_{\newsub{\transys}}(\loc,\mathbf{x}). \]
	Next, define a map $\oldsub{\potential}$ that to each reachable state $(\loc,\mathbf{x})$ in $\oldsub{\transys}$ assigns
	\[ \oldsub{\antipotential}(\loc,\mathbf{x}) := \CostSup_{\oldsub{\transys}}(\loc,\mathbf{x}). \]
	Then, the same argument as in the proof of Theorem~\ref{thm:potentialrelational} shows that $\newsub{\antipotential}$ is an anti-PF in $\newsub{\transys}$ and that $\oldsub{\potential}$ is a PF in $\oldsub{\transys}$. Moreover, since both transition systems are deterministic, we have $\CostInf_{\newsub{\transys}}(\loc,\mathbf{x}) = \CostSup_{\newsub{\transys}}(\loc,\mathbf{x})$ for each reachable state $(\loc,\mathbf{x})$ in $\newsub{\transys}$ and analogously for $\oldsub{\transys}$. Hence, as $t$ is not a threshold for the DiffCost problem, we conclude that there exists $\mathbf{x}\in\initvars$ such that
	\begin{equation*}
	\begin{split}
	t &< \CostSup_{\newsub{\transys}}(\newsub{\loc_0},\mathbf{x}) - \CostInf_{\oldsub{\transys}}(\oldsub{\loc_0},\mathbf{x}) \\ 
	&= \CostInf_{\newsub{\transys}}(\newsub{\loc_0},\mathbf{x}) - \CostSup_{\oldsub{\transys}}(\oldsub{\loc_0},\mathbf{x}) \\
	&= \newsub{\antipotential}(\newsub{\locinit},\mathbf{x}) - \oldsub{\potential}(\oldsub{\locinit},\mathbf{x}).
	\end{split}
	\end{equation*}
	So $\newsub{\antipotential}$ and $\oldsub{\potential}$ satisfy the theorem claim.
\end{proof}

\begin{manualtheorem}{5.1}[Soundness]
	If the algorithm outputs a value $t$, then $t$ is a threshold for the DiffCost problem. Furthermore, the algorithm runs in polynomial time.
\end{manualtheorem}

\begin{proof}
	To establish soundness of our algorithm, we need to show that:
	\begin{enumerate}
		\item Every solution to the system of constraints produced in Step~2 gives rise to a theshold value $t$ and a pair of a PF $\newsub{\potential}$ and an anti-PF $\oldsub{\antipotential}$ that witness it.
		\item Every solution to the system of constraints produced in Step~3 gives rise to a theshold value $t$ and a pair of a PF $\newsub{\potential}$ and an anti-PF $\oldsub{\antipotential}$ that witness it.
	\end{enumerate}

	The first item is true since the defining properties of the PF $\newsub{\potential}$ (the Sufficiency preservation and the Sufficiency on termination conditions) and the defining properties of the anti-PF $\oldsub{\antipotential}$ (the Insufficiency preservation and the Insufficiency on termination conditions) are imposed at all states contained in $\newsub{I}$ and $\oldsub{I}$, respectively. Since $\newsub{I}$ and $\oldsub{I}$ over-approximate the sets of states reachable in $\newsub{\transys}$ and $\oldsub{\transys}$, it follows that any solution to the system of constraints produced in Step~2 gives rise to a threshold value $t$ and a pair of a PF $\newsub{\potential}$ and an anti-PF $\oldsub{\antipotential}$ that witness it.
	
	The second item is true due to the soundness of translation of the constraint in eq.~\eqref{eq:constraint} into the constraint in eq.~\eqref{eq:handelmanaffine} which was established in Section~\ref{sec:algorithm}.
	
	The fact that the algorithm runs in polynomial time follows since Step~1, Step~2 and Step~3 all take polynomial time in the size of the programs (when parametrized by the maximal polynomial degree $d$). Moreover, the size of the system of linear constraints $\Phi$ produced in Step~3 is polynomial in the size of the programs. Since linear programming instances can be solved in polynomial time, we conclude that our algorithm runs in time polynomial in the size of the programs.
\end{proof}

\begin{manualtheorem}{7.1}
	Let $\transys$ be a terminating transition systems. Suppose that $\potential$ is a PF and that $\antipotential$ is an anti-PF in $\transys$. Then, for each initial variable valuation $\mathbf{x}\in\initvars$, we have that
	\begin{equation*}
	\CostSup_{\transys}(\locinit,\mathbf{x}) - \CostInf_{\transys}(\locinit,\mathbf{x}) \leq \potential(\locinit,\mathbf{x}) - \antipotential(\locinit,\mathbf{x}).
	\end{equation*}
	In particular, if $p$ satisfies $\potential(\locinit,\mathbf{x}) - \antipotential(\locinit,\mathbf{x}) \leq p$ for each $\mathbf{x}\in\initvars$, then for any run $\rho$ that starts in some initial state $(\locinit,\mathbf{x})$ with $\mathbf{x}\in\Theta_0$ we have $ 0\leq \Cost_{\transys}(\rho)-\antipotential(\locinit,\mathbf{x}) \leq p$ and $ 0\leq \potential(\locinit,\mathbf{x}) - \Cost_{\transys}(\rho) \leq p$. Hence, $p$ is a bound on the precision of the upper cost bound defined by $\potential$ and the lower cost bound defined by $\antipotential$ in $\transys$.
\end{manualtheorem}

\begin{proof}
	By Theorem~\ref{thm:potential}, it follows that
	\[ \CostSup_{\transys}(\loc_0,\mathbf{x}) \leq \potential(\locinit,\mathbf{x}) \]
	and
	\[ \CostInf_{\transys}(\loc_0,\mathbf{x}) \geq \antipotential(\locinit,\mathbf{x}). \]
	Combining the two inequalities, we conclude that
	\begin{equation*}
	\begin{split}
	&\CostSup_{\transys}(\loc_0,\mathbf{x}) - \CostInf_{\transys}(\loc_0,\mathbf{x}) \\
	&\leq \potential(\locinit,\mathbf{x}) - \antipotential(\locinit,\mathbf{x}),
	\end{split}
	\end{equation*}
	as desired. Since for each run $\rho$ that starts in some initial state $(\locinit,\mathbf{x})$ with $\mathbf{x}\in\Theta_0$ we have
	\begin{equation*}
	\begin{split}
	\antipotential(\loc_0,\mathbf{x}) \leq \CostInf_{\transys}(\loc_0,\mathbf{x}) &\leq \Cost_{\transys}(\rho) \\
	&\leq \CostSup_{\transys}(\loc_0,\mathbf{x}) \leq \potential(\loc_0,\mathbf{x})
	\end{split}
	\end{equation*}
	if $p$ satisfies $\potential(\locinit,\mathbf{x}) - \antipotential(\locinit,\mathbf{x}) \leq p$ for each $\mathbf{x}\in\initvars$ then it follows that $ 0\leq \Cost_{\transys}(\rho)-\antipotential(\locinit,\mathbf{x}) \leq p$ and $ 0\leq \potential(\locinit,\mathbf{x}) - \Cost_{\transys}(\rho) \leq p$.
\end{proof}

\section{Necessity of Termination Assumption in Theorem~\ref{thm:potential}}\label{app:nonterm}

To show the necessity of the termination assumption in Theorem~\ref{thm:potential}, consider the program

\begin{lstlisting}[mathescape]
      void nonterm(int x){
            int cost = 0;
$\locinit:$	 while (x >= 0) { 
$\loc_1:$             if (x <= 5) {
$\loc_2:$	          cost = cost + 1;
                }
$\loc_3:$             x = x + 1; 
            }
      }
$\locterm:$
\end{lstlisting}

\noindent and define $\antipotential$ as follows:
\begin{equation*}
\antipotential(\loc,x,\cost) = \begin{cases}
7-x, &\text{if } \loc\in\{\locinit,\loc_1,\loc_2\} \text{ and } 0\leq x\leq 5 \\
6-x, &\text{if } \loc=\loc_3 \text{ and } 0\leq x\leq 5 \\
1, &\text{otherwise}
\end{cases}
\end{equation*}
One can verify by inspection that $\antipotential$ is an antipotential function for this program (note that the insufficiency on termination condition is trivial as the program does not terminate). However, $\antipotential(\locinit,0,0) = 7$ exceeds the total cost usage, which is equal to $6$. Hence, $\antipotential$ would contradict the claim for anti-PFs in Theorem~\ref{thm:potential} and the termination assumption in Theorem~\ref{thm:potential} is necessary for the theorem claim to be correct.
